\documentclass[aps,twocolumn,showpacs,superscriptaddress]{revtex4-1}
\usepackage[T1]{fontenc}
\usepackage{amsfonts,amsmath,amssymb,amsbsy}
\usepackage{graphicx,color}
\usepackage{times}
\usepackage[colorlinks=true, linkcolor=blue, citecolor=red, urlcolor=magenta]{hyperref}

\let\mathbf=\boldsymbol
\def\D{{\mathcal{D}}}
\def\B{{\mathcal{B}}}
\def\I{{\mathcal{I}}}

\def\emph#1{\textcolor{red}{#1}}
\begin{document}

\title{Control and manipulation of a magnetic skyrmionium in nanostructures}

\author{Xichao Zhang}
\affiliation{School of Science and Engineering, The Chinese University of Hong Kong, Shenzhen 518172, China}

\author{Jing Xia}
\affiliation{School of Science and Engineering, The Chinese University of Hong Kong, Shenzhen 518172, China}

\author{Yan Zhou}
\email[]{zhouyan@cuhk.edu.cn}
\affiliation{School of Science and Engineering, The Chinese University of Hong Kong, Shenzhen 518172, China}

\author{Daowei Wang}
\affiliation{School of Physics and Electronics, Central South University, Changsha 410083, China}

\author{Xiaoxi Liu}
\affiliation{Department of Information Engineering, Shinshu University, Wakasato 4-17-1, Nagano 380-8553, Japan}

\author{Weisheng Zhao}
\affiliation{Fert Beijing Institute, Beihang University, Beijing 100191, China}
\affiliation{School of Electronic and Information Engineering, Beihang University, Beijing 100191, China}

\author{Motohiko Ezawa}
\email[]{ezawa@ap.t.u-tokyo.ac.jp}
\affiliation{Department of Applied Physics, University of Tokyo, Hongo 7-3-1, Tokyo 113-8656, Japan}

\begin{abstract}
A magnetic skyrmionium is a nontopological soliton, which has a doughnut-like out-of-plane spin texture in thin films, and can be phenomenologically viewed as a coalition of two topological magnetic skyrmions with opposite topological numbers. Due to its zero topological number ($Q=0$) and doughnut-like structure, the skyrmionium has its distinctive characteristics as compared to the skyrmion with $Q=\pm 1$. Here we systematically study the generation, manipulation and motion of a skyrmionium in ultrathin magnetic nanostructures by applying a magnetic field or a spin-polarized current. It is found that the skyrmionium moves faster than the skyrmion when they are driven by the out-of-plane current, and their velocity difference is proportional to the driving force. However, the skyrmionium and skyrmion exhibit an identical current-velocity relation when they are driven by the in-plane current. It is also found that a moving skyrmionium is less deformed in the current-in-plane geometry compared with the skyrmionum in the current-perpendicular-to-plane geometry. Furthermore we demonstrate the transformation of a skyrmionium with $Q=0$ into two skyrmions with $Q=+1$ in a nanotrack driven by a spin-polarized current, which can be seen as the unzipping process of a skyrmionium. We illustrate the energy and spin structure variations during the skyrmionium unzipping process, where linear relations between the spin structure and energies are found. These results could have technological implications in the emerging field of skyrmionics.
\end{abstract}

\date{\today}
\keywords{magnetic skyrmion, magnetic skyrmionium, skyrmion Hall effect, spintronics}
\pacs{75.60.Ch, 75.70.Kw, 75.78.Cd, 12.39.Dc}

\maketitle

\section{Introduction}
\label{se:Introduction}

Magnetic skyrmion has emerged as an active research topic in the recent decade, as it is a stable topological soliton anticipated to be a fundamental component in future magnetic data storage and spintronics applications~\cite{Roszler_NATURE2006,Nagaosa_NNANO2013,Seki_BOOK2016}. Recent experiments have confirmed the existence of the skyrmion in a wide variety of magnetic materials and structures~\cite{Muhlbauer_SCIENCE2009,Yu_NATURE2010,Heinze_NPHYS2011,Schulz_NPHYS2012,Romming_SCIENCE2013,Finazzi_PRL2013,Kezsmarki_NMATER2015,Schwarze_NMATER2015,Du_NCOMMS2015,Nii_NCOMMS2015,Wanjun_SCIENCE2015,Boulle_NNANO2016}. Meanwhile, numerous theoretical and numerical studies have revealed various potential applications of skyrmions toward the skyrmion-electronics as well as the skyrmion-spintronics~\cite{Iwasaki_NC, Iwasaki_NL,Sampaio_NNANO2013,Iwasaki_NNANO2013,Sun_PRL2013,Tomasello_SREP2014,Yan_NCOMMS2014,Xichao_SREP2015A,Xichao_SREP2015B,Yan_NCOMMS2015,Upadhyaya_PRB2015,Koshibae_JJAP2015}.

However, there is a significant obstacle to the transmission of skyrmions in confined geometries, especially in the high-speed operation, that is, the so-called skyrmion Hall effect (SkHE). The SkHE, which was previously theoretically predicted~\cite{Zang_PRL2001} and has recently been observed experimentally~\cite{Wanjun_ARXIV2016}, is caused by the Magnus force acting on the moving skyrmion with a topological number of $Q=\pm 1$. The SkHE is generally a detrimental effect since the skyrmion experiencing it will deviate from the desired transmission path, which may ultimately lead to the destruction of the skyrmion at the edge of the device. In light of this, strategies are urgently required to overcome the SkHE. One promising approach is to construct an antiferromagnetically exchange-coupled bilayer system~\cite{Xichao_NCOMMS2016,Xichao_ARXIV2016}, in which the bilayer-skyrmion is a combination of a skyrmion with $Q=+1$ in the front layer and a skyrmion with $Q=-1$ in the back layer. It is an object with $Q=0$, and thus totally free of the SkHE.

\begin{figure}[t]
\centerline{\includegraphics[width=0.50\textwidth]{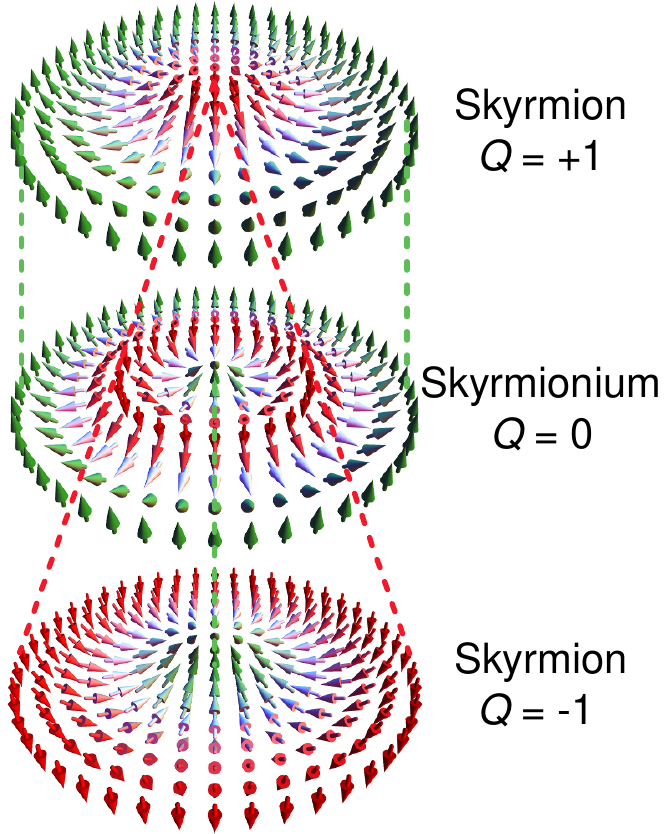}}
\caption{(Color online)
Illustrations of a magnetic skyrmionium with a topological number of $Q=0$, which can be seen as composed of a skyrmion with $Q=+1$ and a skyrmion with $Q=-1$. The arrows denote the spin directions. The dashed lines indicate the spin-to-spin mapping relation among the skyrmion with $Q=+1$, the skyrmion with $Q=-1$, and the skyrmionium with $Q=0$.
}
\label{FIG1}
\end{figure}

In this respect, the magnetic skyrmionium is an intriguing object in monolayer thin films, because it has a skyrmion-like structure and yet has a topological number of $Q=0$~\cite{Bogdanov_JMMM1999}. It is a composite structure made of a skyrmion with $Q=+1$ and a skyrmion with $Q=-1$ forming a doughnut-like out-of-plane spin distribution~\cite{Bogdanov_JMMM1999,Rohart_PRB2013,Liu_AIPADV2015,Beg_SREP2015,Komineas_PRB2015,Komineas_ARXIV2015,Beg_ARXIV2016,Leonov_ARXIV2013,Liu_PRB2015}, as illustrated in Fig.~\ref{FIG1}. It has been theoretically suggested that the skyrmionium can be created and remain stable in magnetic nanodisks with the Dzyaloshinskii-Moriya interaction (DMI)~\cite{Bogdanov_JMMM1999,Leonov_ARXIV2013,Rohart_PRB2013,Liu_AIPADV2015,Beg_SREP2015,Liu_PRB2015,Mulkers_PRB2016}. Indeed, the first experimental observation of the skyrmionium has been achieved on a ferromagnetic (FM) thin film by laser radiation with a higher energy as compare to the skyrmion generation~\cite{Finazzi_PRL2013}, where each skyrmionium thus created has been found stable over a few years. More recently, and significantly, it was predicted theoretically that the skyrmionium can be driven into motion by the magnetic field gradient as well as the in-plane spin-polarized current~\cite{Komineas_PRB2015,Komineas_ARXIV2015}, indicating its importance in future electronic and spintronic applications.

In spite of its importance, the skyrmionium has not been much studied and many of its properties remain elusive. In this paper, we show that the skyrmionium has several distinctive features such as different generation, destruction and transportation behaviors as compared to the skyrmion. In particular we study carefully the motion dynamics of a skyrmionium in the nanotrack induced by either the in-plane or out-of-plane spin-polarized current injection. It is found that the steady velocity difference between the skyrmionium and skyrmion is proportional to the driving current density when they are driven by the out-of-plane current in the nanotrack. However, the skyrmionium and skyrmion have no steady velocity difference when they are driven by the in-plane current in the nanotrack. The moving skyrmionium has the trend to be distorted since the skyrmion with $Q=+1$ and the skyrmion with $Q=-1$ components are affected by the SkHEs working in opposite directions. Importantly, we find that the distortion of the skyrmionium driven by the out-of-plane current is much more significant than the skyrmionium driven by the in-plane current with the same current density. Nevertheless, intriguingly there are parameter regions where the skyrmionium travels straightforward without deformation or obvious deformation. Furthermore, we demonstrate that it is possible to degenerate a skyrmionium with $Q=0$ into a skyrmion with $Q=+1$ or $Q=-1$ in a nanodisk by applying a magnetic field. We also illustrate the transformation of a skyrmionium with $Q=0$ into two skyrmions with $Q=+1$ in a nanotrack driven by a spin-polarized current, which can be seen as the unzipping process of a skyrmionium. We investigate the energy and spin structure variations during the unzipping process of a skyrmionium, where linear relations between the spin structures and energies are found. These results provide a base for utilizing nontopological skyrmioniums for designing novel information processing, storage and logic computing devices.

\section{Magnetization Dynamics Equations}
\label{se:Methods}

In our simulation, we model the ultrathin magnetic nanodisk with the radius $r$, as well as the ultrathin magnetic nanotrack with the length $l$ and the width $w$, where the thickness of the FM layer in all models is fixed at $a=1$ nm throughout the paper.

We perform the three-dimensional (3D) micromagnetic simulation with the 1.2 alpha 5 release of the Object Oriented MicroMagnetic Framework (OOMMF) software developed at the National Institute of Standards and Technology (NIST)~\cite{OOMMF}. The simulation is carried out by a set of the OOMMF extensible solver (OXS) objects of the standard OOMMF distribution. Indeed, we include the OXS extension module~\cite{Rohart_PRB2013} in order to model the interface-induced DMI in our simulation. In the absence of the spin-polarized current, the 3D magnetization dynamics in the FM layer is governed by the standard Landau-Lifshitz-Gilbert (LLG) equation~\cite{Gilbert_LLG,Landau_LLG,OOMMF}
\begin{equation}
\frac{d\boldsymbol{M}}{dt}=-\gamma_{\text{0}}\boldsymbol{M}\times\boldsymbol{H}_{\text{eff}}+\frac{\alpha}{M_{\text{S}}}(\boldsymbol{M}\times\frac{d\boldsymbol{M}}{dt}),
\label{eq:LLG}
\end{equation}
where $\boldsymbol{M}$ is the magnetization, $M_{\text{S}}=|\boldsymbol{M}|$ is the saturation magnetization, $t$ is the time, $\gamma_{\text{0}}$ is the gyromagnetic ratio with absolute value, and $\alpha$ is the Gilbert damping coefficient. $\boldsymbol{H}_{\text{eff}}$ is the effective field, which reads
\begin{equation}
\boldsymbol{H}_{\text{eff}}=-\mu_{0}^{-1}\frac{\partial E}{\partial \boldsymbol{M}}.
\label{eq:H-eff}
\end{equation}
The average energy density $E$ contains the Heisenberg exchange, the perpendicular magnetic anisotropy (PMA), the applied magnetic field, the demagnetization, and the DMI energy terms, which is expressed as follows
\begin{align}
\label{eq:E} 
E&=A[\nabla(\frac{\boldsymbol{M}}{M_{\text{S}}})]^{2}-K\frac{(\boldsymbol{n}\cdot\boldsymbol{M})^{2}}{M_{\text{S}}^{2}} \\ \notag
&-\mu_{0}\boldsymbol{M}\cdot\boldsymbol{H}-\frac{\mu_{0}}{2}\boldsymbol{M}\cdot\boldsymbol{H}_{\text{d}}(\boldsymbol{M}) \\ \notag
&+\frac{D}{M_{\text{S}}^{2}}(M_{z}\frac{\partial M_{x}}{\partial x}+M_{z}\frac{\partial M_{y}}{\partial y}-M_{x}\frac{\partial M_{z}}{\partial x}-M_{y}\frac{\partial M_{z}}{\partial y}), 
\end{align}
where $A$, $K$, and $D$ are the Heisenberg exchange, PMA, and DMI energy constants, respectively. $\boldsymbol{H}$ is the applied magnetic field, and $\boldsymbol{H}_{\text{d}}(\boldsymbol{M})$ is the demagnetization field. The $M_x$, $M_y$ and $M_z$ are the three Cartesian components of the magnetization $\boldsymbol{M}$. $\mu_0$ is the vacuum permeability constant, and $\boldsymbol{n}$ is the unit surface normal vector. When the effect of the spin-polarized current is taken into account, the 3D magnetization dynamics in the FM layer is governed by the extended LLG equation including the spin transfer torques (STTs). For the current-perpendicular-to-plane (CPP) geometry, the LLG equation (\ref{eq:LLG}) is extended into the form
\begin{align}
\label{eq:LLGS-CPP}
\frac{d\boldsymbol{M}}{dt}=&-\gamma_{0}\boldsymbol{M}\times\boldsymbol{H}_{\text{eff}}+\frac{\alpha}{M_{\text{S}}}(\boldsymbol{M}\times\frac{d\boldsymbol{M}}{dt}) \\ \notag
&+\frac{u}{aM_{\text{S}}}(\boldsymbol{M}\times \boldsymbol{p}\times \boldsymbol{M}),
\end{align}
where $u=|\frac{\gamma_{0}\hbar}{\mu_{0}e}|\frac{jP}{2M_{\text{S}}}$ is the STT coefficient, and $\boldsymbol{p}$ stands for the unit spin polarization direction. $\hbar$ is the reduced Planck constant, $e$ is the electron charge, $j$ is the applied current density, and $P$ is the spin polarization rate. We set $\boldsymbol{p}=-\hat{z}$ for creating the skyrmionium, and $\boldsymbol{p}=-\hat{y}$ for driving the skyrmionium or the skyrmion. For the current-in-plane (CIP) geometry, the LLG equation (\ref{eq:LLG}) is extended into the form
\begin{align}
\label{eq:LLGS-CIP}
\frac{d\boldsymbol{M}}{dt}=&-\gamma_{0}\boldsymbol{M}\times\boldsymbol{H}_{\text{eff}}+\frac{\alpha}{M_{\text{S}}}(\boldsymbol{M}\times \frac{d\boldsymbol{M}}{dt}) \\ \notag
&+\frac{u}{M_{\text{S}}^2}(\boldsymbol{M}\times\frac{\partial\boldsymbol{M}}{\partial x}\times\boldsymbol{M})-\frac{\beta u}{M_{\text{S}}}(\boldsymbol{M}\times\frac{\partial\boldsymbol{M}}{\partial x}),
\end{align}
with $\beta$ being the strength of the nonadiabatic STT torque. The intrinsic magnetic material parameters used in our simulation are adopted from Refs.~\onlinecite{Sampaio_NNANO2013,Yan_NCOMMS2014,Yan_NCOMMS2015}: $\gamma_{0}=2.211\times 10^{5}$ m A$^{-1}$ s$^{-1}$, $\alpha=0.3$, $M_{\text{S}}=580$ kA m$^{-1}$, $A=15$ pJ m$^{-1}$, $K=0.8$ MJ m$^{-3}$, $D=0\sim 6$ mJ m$^{-2}$, and $P=0.4$. All models are discretized into tetragonal elements with the size of $2a$ $\times$ $2a$ $\times$ $a$, which ensures a good compromise between computational accuracy and efficiency.

\begin{figure}[t]
\centerline{\includegraphics[width=0.50\textwidth]{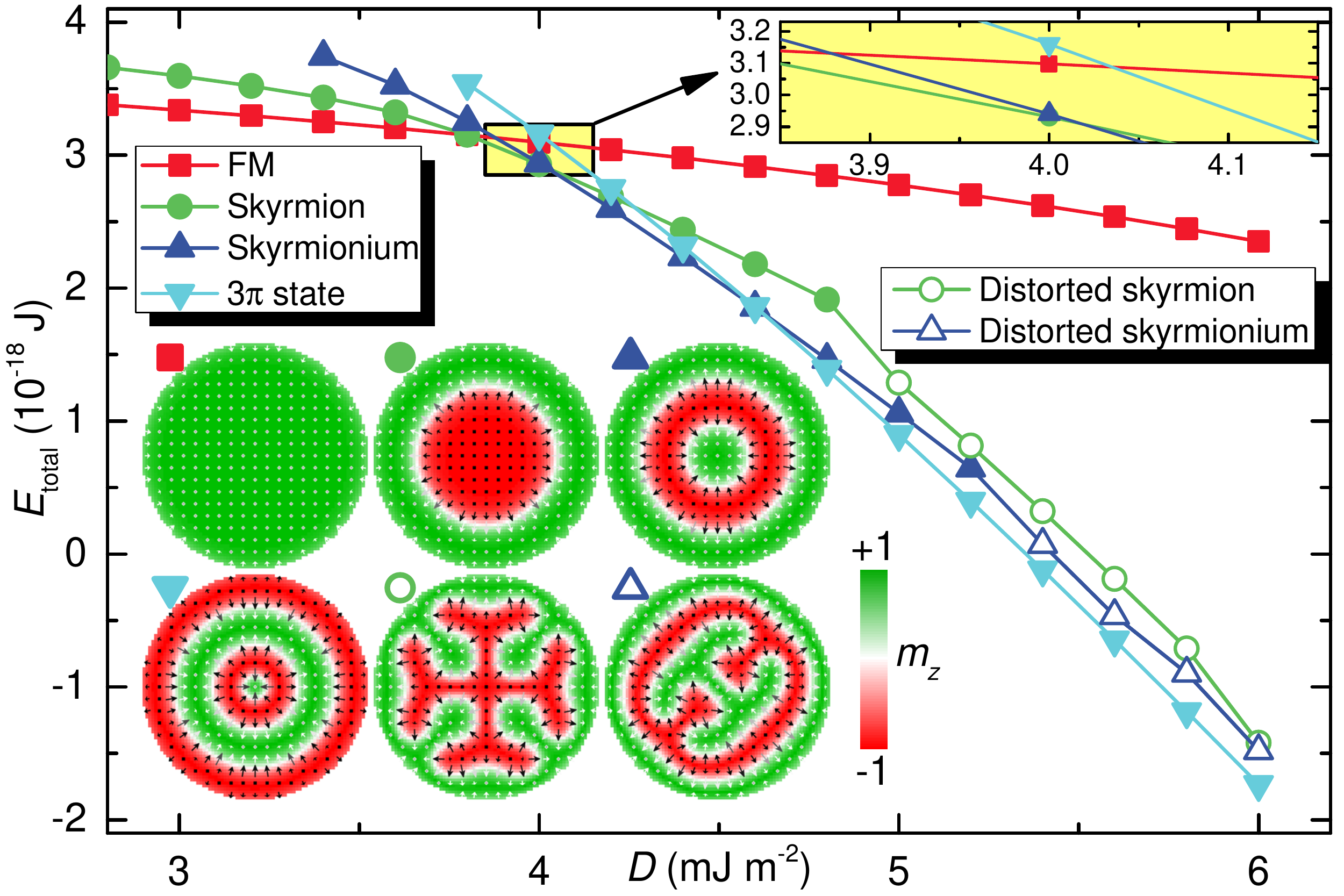}}
\caption{(Color online)
Total micromagnetic energy $E_{\text{total}}$ for FM state, skyrmion, skyrmionium, and $3\pi$ state as a function of the DMI constant $D$ in the nanodisk with $r=75$ nm. $E_{\text{total}}$ includes the Heisenberg exchange, PMA, demagnetization, and DMI energies. A zoomed view of the curves around $D=4$ mJ m$^{-2}$ is given. Insets show the top-views of the FM state, skyrmion, skyrmionium, and $3\pi$ state at $D=4$ mJ m$^{-2}$, and that of the distorted skyrmion and skyrmionium at $D=5.6$ mJ m$^{-2}$. The color scale, which has been used throughout the paper, represents the out-of-plane magnetization component $m_z$.
}
\label{FIG2}
\end{figure}

On the other hand, both the skyrmionium and skyrmion are characterized by the topological number $Q$, that is, the Pontryagin number, which is given as
\begin{equation}
Q=-\frac{1}{4\pi}\int\boldsymbol{m}\cdot(\frac{\partial\boldsymbol{m}}{\partial x}\times\frac{\partial\boldsymbol{m}}{\partial y})dxdy,
\label{eq:Q}
\end{equation}
with $\boldsymbol{m}=\boldsymbol{M}/M_{\text{S}}$ being the normalized magnetization. The topological number $Q$ is also referred to as the skyrmion number. A skyrmion has a topological number of $Q=\pm 1$, while a skyrmionium has a topological number of $Q=0$. It should be noted that the topological number $Q$ takes an integer number only in a continuous and infinitely large system without boundary. As our system has a underlying lattice structure and a finite size, the topological number $Q$ is not quantized. We may even create or destroy a skyrmion in such a system. See Ref.~\onlinecite{Xichao_PRB2016} for a detailed mechanism of these processes in an instance of the topological charge $Q=2$. Nevertheless, a skyrmion is topologically stable as far as it is far away from the boundary, where $Q$ is almost quantized.

\section{Generation of a skyrmionium}
\label{se:Generation}

\begin{figure}[t]
\centerline{\includegraphics[width=0.50\textwidth]{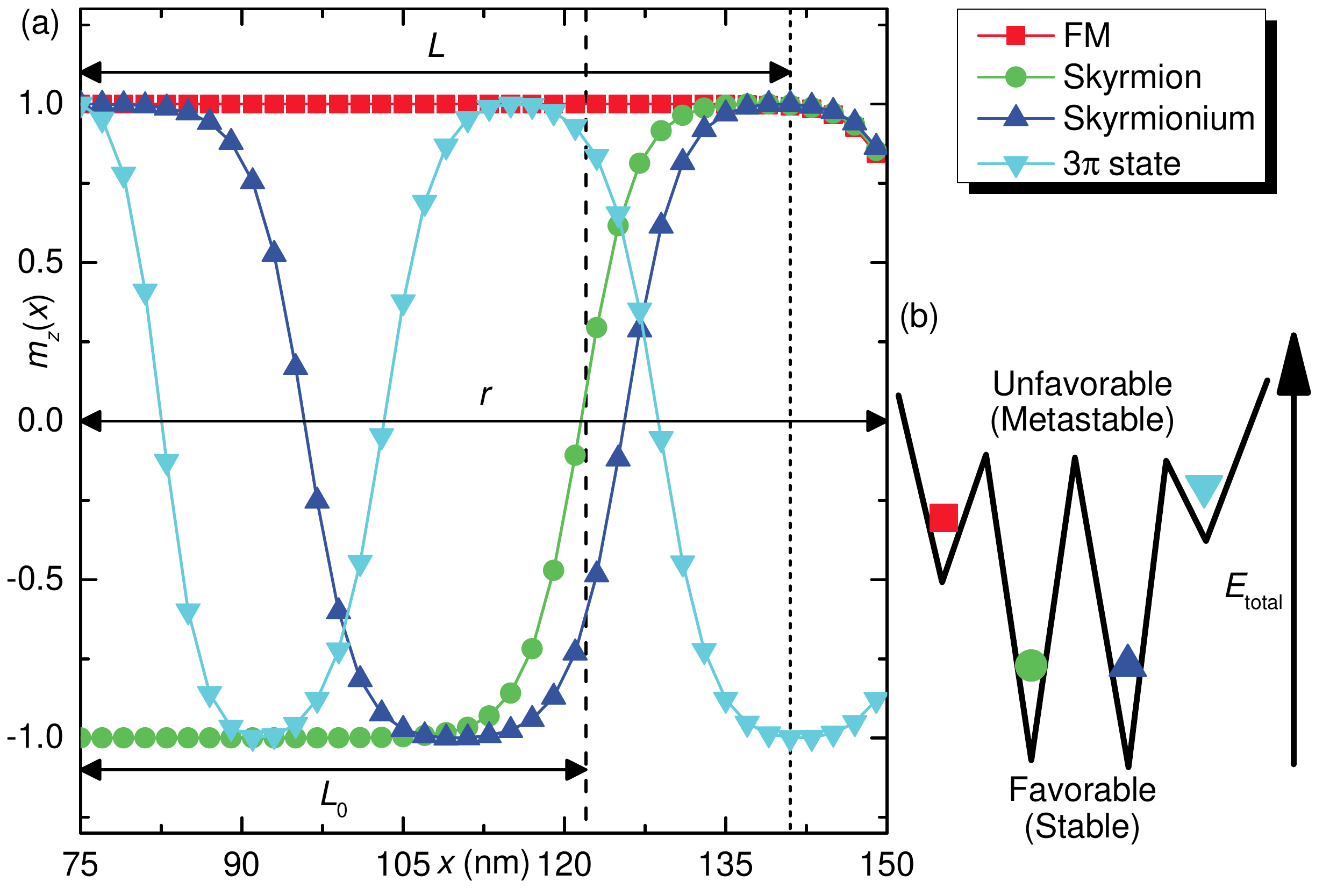}}
\caption{(Color online)
(a) Relaxed profiles $m_z(x)$ of the out-of-plane magnetization component along the radius for the FM state, skyrmion, skyrmionium, and $3\pi$ state in the nanodisk with $r=75$ nm and $D=4$ mJ m$^{-2}$. $L=66$ nm and $L_0=47$ nm stand for the cycloid period of the skyrmionium and the anisotropy-free cycloid period, respectively.
(b) Sketch of the four-well potential of the nanodisk with $r=75$ nm and $D=4$ mJ m$^{-2}$, where the skyrmion and skyrmionium are favorable (stable) states, while the FM state and $3\pi$ state are unfavorable (metastable).
}
\label{FIG3}
\end{figure}

\begin{figure*}[t]
\centerline{\includegraphics[width=1.00\textwidth]{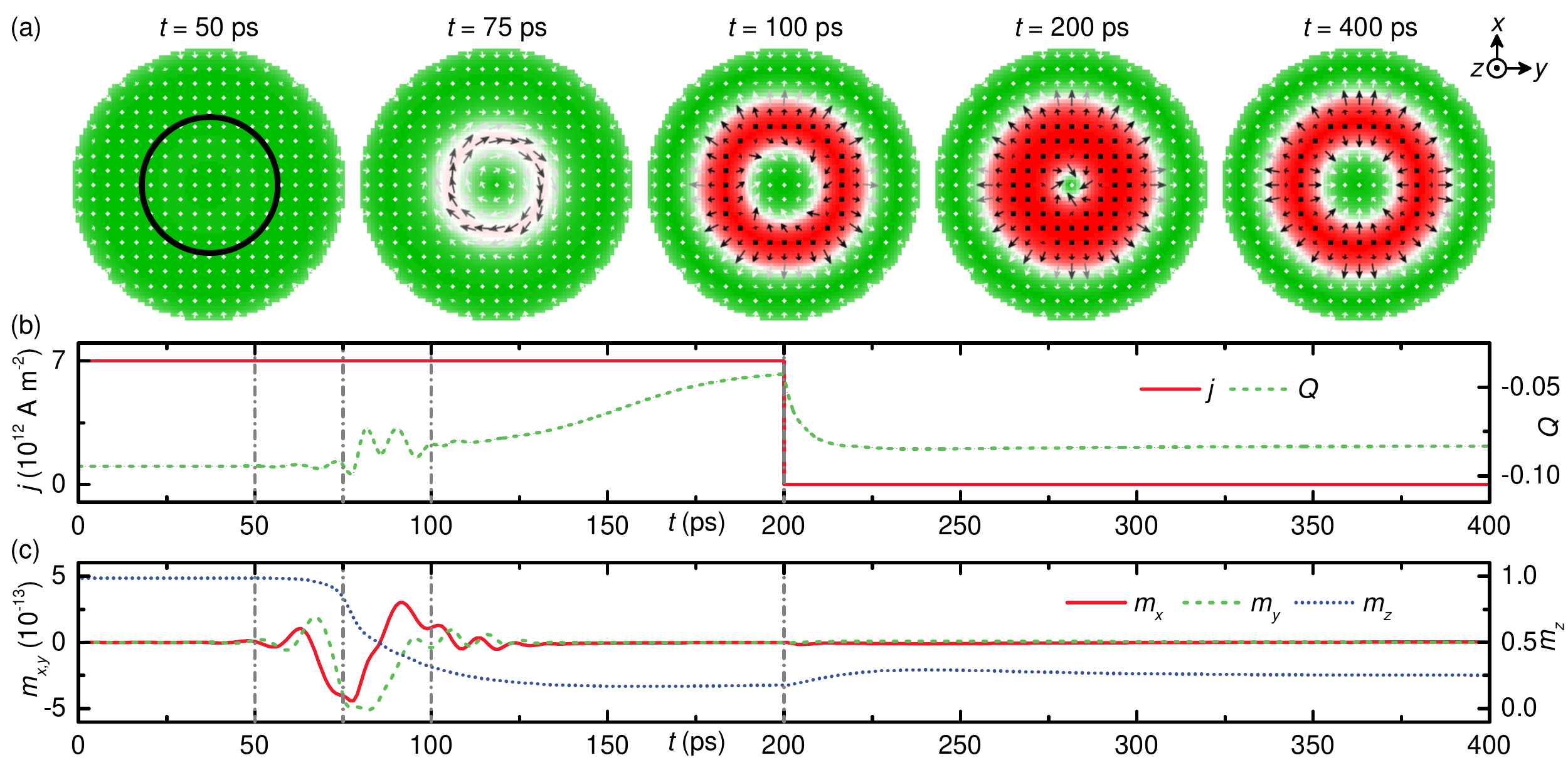}}
\caption{(Color online)
Generation of a skyrmionium by applying a spin-polarized current with the CPP geometry in the nanodisk with $r=75$ nm and $D=4$ mJ m$^{-2}$.
(a) Top-views of the magnetization distribution at selected $t$. A $200$-ps-long spin-polarized current is injected into the central region with a radius of $r_{\text{i}}=r/2$, which is indicated by the black circle.
(b) The spin-polarized current density $j$ and the topological number $Q$ as functions of $t$.
(c) The normalized in-plane and out-of-plane magnetization components ($m_x$, $m_y$, $m_z$) as functions of $t$.
}
\label{FIG4}
\end{figure*}

\begin{figure*}[t]
\centerline{\includegraphics[width=1.00\textwidth]{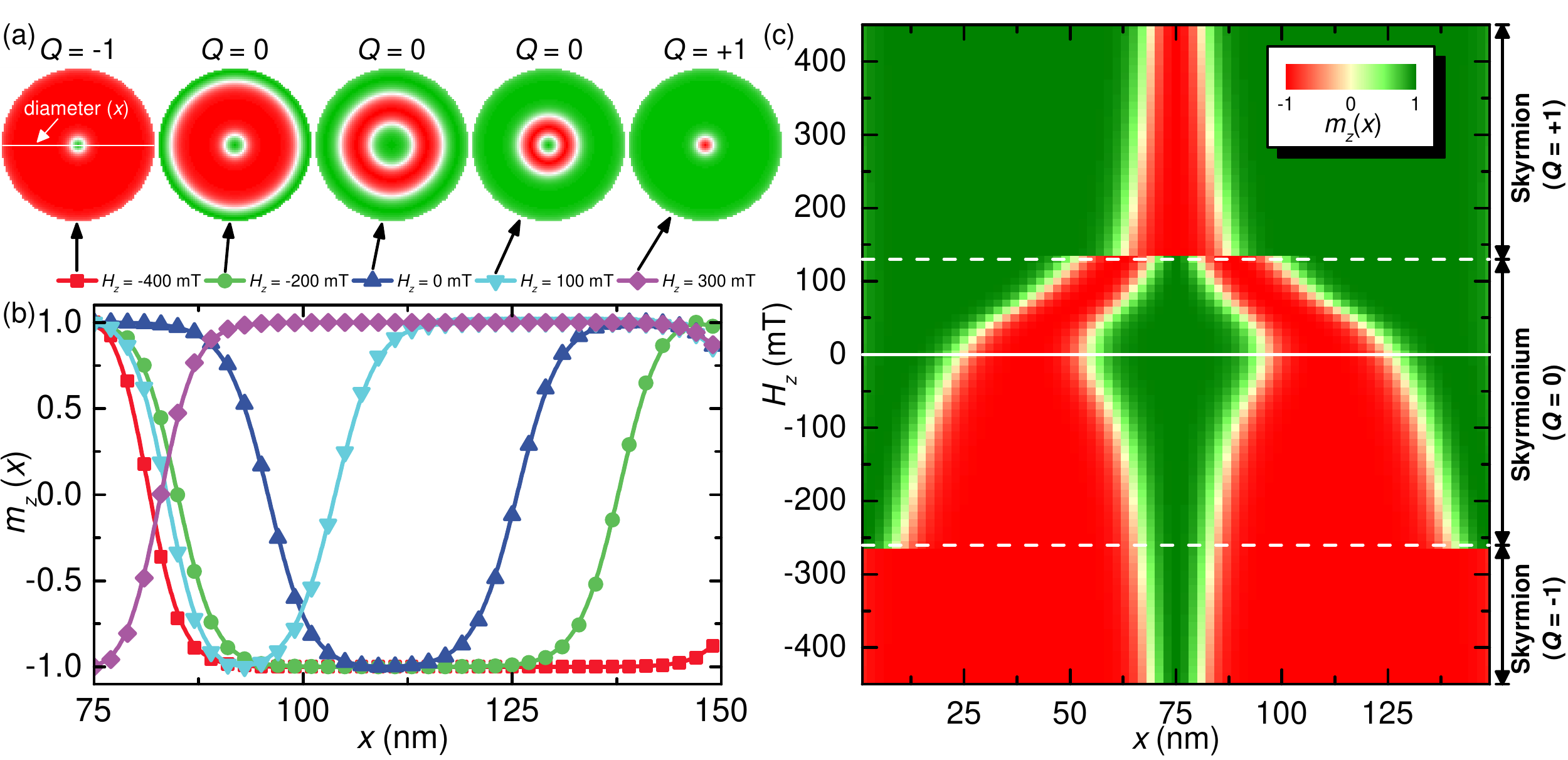}}
\caption{(Color online)
Dependence of the size and shape of the skyrmionium on the applied perpendicular magnetic field $H_{z}$ in the nanodisk with $r=75$ nm and $D=4$ mJ m$^{-2}$.
(a) Top-views of the magnetization distribution at selected $H_{z}$.
(b) Relaxed profile $m_z(x)$ along the nanodisk radius at selected $H_{z}$.
(c) Relaxed profile $m_z(x)$ along the nanodisk diameter as a function of $H_{z}$. $H_{z}>0$ ($H_{z}<0$) denotes the field is applied along the $+z$-direction ($-z$-direction). When $H_{z}>130$ mT or $H_{z}<-260$ mT, the degeneration of a skyrmionium with $Q=0$ to a skyrmion with $Q=\pm 1$ occurs.
}
\label{FIG5}
\end{figure*}

First, we compare the energy level of the skyrmionium with possible states, that is, the FM state, the skyrmion, and the $3\pi$ rotation state, in a magnetic nanodisk with the DMI. The radius of the nanodisk is fixed at $r=75$ nm. By giving the the FM state, skyrmion, skyrmionium, and $3\pi$ state as the initial magnetization configuration of the nanodisk separately, we calculate the total micromagnetic energy $E_{\text{total}}$ of the relaxed nanodisk as a function of the DMI constant $D$. The results are given in Fig.~\ref{FIG2}.

It can be seen that the FM state is the most stable state when $D<3.8$ mJ m$^{-2}$, while the skyrmion and skyrmionium become more stable than the FM state when $D>3.8$ mJ m$^{-2}$. When $D>4$ mJ m$^{-2}$, the $3\pi$ state also becomes more stable than the FM state. Hence, we identify the $D_{\text{c}}=3.8$ mJ m$^{-2}$ as the critical DMI constant (see Ref.~\onlinecite{Rohart_PRB2013}) in the given nanodisk, beyond which the domain walls correspond to an energy gain and the FM state is thus unfavorable metastable or unstable state. At $D=4$ mJ m$^{-2}$, we find that the skyrmion and skyrmionium are favorable stable states as they have almost the same and lowest $E_{\text{total}}$. That is to say, the FM state and $3\pi$ state, having higher $E_{\text{total}}$, are unfavorable metastable states. The magnetization configurations of the four states at $D=4$ mJ m$^{-2}$ are shown in Fig.~\ref{FIG2} inset. When $D>4$ mJ m$^{-2}$, $E_{\text{total}}$ of the skyrmionium is lower than that of the skyrmion. And when $D>4.6$ mJ m$^{-2}$, the $3\pi$ state has the lowest $E_{\text{total}}$ compared to the other three states.

It should be noted that the relaxed skyrmion and the relaxed skyrmionium are spontaneously distorted, when $D\geq 5$ mJ m$^{-2}$ and $D\geq 5.4$ mJ m$^{-2}$, respectively. The magnetization configurations of the distorted skyrmion and the distorted skyrmionium at $D=5.6$ mJ m$^{-2}$ are also shown in Fig.~\ref{FIG2} inset. The distorted skyrmion and the distorted skyrmionium form when $D$ is larger than certain values, as larger $D$ favors the formation of more and longer domain walls~\cite{Rohart_PRB2013}. For example, at the very large $D=5.6$ mJ m$^{-2}$, the $3\pi$ state has the lowest $E_{\text{total}}$ compared to the skyrmion and skyrmionium, indicating three circular domain walls are more favorable than two or one. Therefore, the skyrmion and skyrmionium have to increase the length of their circular domain walls in order to reach the relaxed metastable states, of which $E_{\text{total}}$ are close to the one of the $3\pi$ state. However, as pointed out in Ref.~\onlinecite{Rohart_PRB2013}, the edge of the nanodisk provides a confinement, which limits the change of the circular domain wall quantity. Hence, the circular domain walls of the skyrmion and skyrmionium with increased length longer than the perimeter of the nanodisk must be distorted in the nanodisk.

For the purposes of studying the generation of an undistorted skyrmionium as well as the transformation from the skyrmionium to the skyrmion, we focus on the nanodisk with $D=4$ mJ m$^{-2}$, where the skyrmion and skyrmionium are favorable, while the FM state and the $3\pi$ state are unfavorable. Figure~\ref{FIG3}(a) shows the relaxed profiles $m_z(x)$ of the out-of-plane magnetization component along the right horizontal radius for the FM state, skyrmion, skyrmionium, and $3\pi$ state, and their relative $E_{\text{total}}$ profile is illustrated in Fig.~\ref{FIG3}(b).

As proved in Ref.~\onlinecite{Rohart_PRB2013}, larger $D$ favors higher magnetization rotation in the nanodisk, especially when the nanodisk radius is larger than the anisotropy-free period $L_0=4\pi A/D$, which describes the period of cycloids developed due to the DMI in the absence of the PMA. For our parameter set, we have $L_0=47$ nm at $D=4$ mJ m$^{-2}$, thus it is reasonable to obtain a stable skyrmionium in the nanodisk with $r=1.6L_0$. On the other hand, according to Ref.~\onlinecite{Rohart_PRB2013}, the cycloid period $L$ is longer than the anisotropy-free $L_0$ at $D/D_{\text{c}}= 1$ in the presence of the PMA. Indeed, as shown in Fig.~\ref{FIG3}(a), where $D/D_{\text{c}}\approx 1$, the cycloid period of the skyrmionium equals $L=1.4L_0=66$ nm, justifying the favorable stability of the skyrmionium at the given condition. Besides, it is obvious that the cycloid period of the $3\pi$ state or even higher magnetization rotation state is shorter than $L_0$, indicating the $3\pi$ state and even higher magnetization rotation state are unfavorable at the given condition. However, as mentioned above and pointed out in Ref.~\onlinecite{Rohart_PRB2013}, the nanodisk edge provides a confinement, that is, a topological barrier, thus the $3\pi$ state can exist here as a metastable state once it is formed artificially.

By using the nanodisk with $D=4$ mJ m$^{-2}$, we study the generation of a skyrmionium, where the skyrmionium is a favorable stable state [see Fig.~\ref{FIG3}(b)] without any distortion once it is created. We perform the generation of a skyrmionium in the nanodisk by applying a spin-polarized current locally with the CPP geometry, as shown in Fig.~\ref{FIG4}(a). The initial magnetization configuration of the nanodisk is the FM state, where the magnetization is aligned along the $+z$-direction. The spin-polarized current of $j=7\times 10^{8}$ A cm$^{-2}$ is injected into the nanodisk within the central region with a radius of $r_{\text{i}}=r/2=37.5$ nm for $200$ ps, followed by an $800$-ps-long relaxation without applying any current. The STT provided by the spin-polarized current reverses the magnetization inside the current injection region.

Figure~\ref{FIG4}(a) shows the top-views of the nanotrack at selected times. The $j$ and $Q$ as functions of $t$ in the generation process are given in Fig.~\ref{FIG4}(b). The corresponding normalized in-plane and out-of-plane magnetization components $m_x$, $m_y$, and $m_z$ of the nanodisk as functions of $t$ are shown in Fig.~\ref{FIG4}(c). It can be seen that, within the current injection region, the magnetization in the core region is pointing along the $+z$-direction while the magnetization around the core is flipped into the plane at $t=75$ ps, resulting in the formation of a doughnut-like region of reversed magnetization. At $t=100$ ps, the magnetization inside the doughnut-like region is reversed to the $-z$-direction, where the magnetization configuration of a skyrmionium is almost developed. The magnetization in the nanodisk reaches the relaxed state at $t=400$ ps, where a stable static skyrmionium is successfully generated.

\section{Manipulation of a skyrmionium}
\label{se:Manipulation}

We have demonstrated the generation of a skyrmionium in the nanodisk by applying a spin-polarized current with the CPP geometry in Sec.~\ref{se:Generation}. Here we continue to investigate the manipulation of a skyrmionium in the nanodisk by applying an external static magnetic field perpendicular to the plane of the nanodisk. The geometry and parameters of the nanodisk are the same as those used for creating a skyrmionium in Sec.~\ref{se:Generation}. A skyrmionium is first created and relaxed in the nanodisk in the absence of the applied magnetic field, that is, $H_{z}=0$ mT, then a magnetic field is perpendicularly applied to the nanodisk with a constant amplitude, leading to a new relaxed state of the system.

Figure~\ref{FIG5}(a) shows the top-views of the relaxed nanodisk at selected values of $H_{z}$, of which the corresponding out-of-plane magnetization component profiles along the right horizontal radius of the nanodisk are given in Fig.~\ref{FIG5}(b). It can be seen that the applied magnetic field within a certain range can adjust the size and shape of the skyrmionium. When the magnetic field is applied perpendicularly to the nanodisk plane and along the $+z$-direction, that is, $H_{z}>0$ mT, the size of the skyrmionium shrinks, where both radiuses of the inner and outer circular domain walls decrease. When the amplitude of the magnetic field is increased to be larger than a certain threshold, that is, $H_{z}=130$ mT, the inner circular domain wall will be annihilated as it shrinks to the lattice size, resulting in the transformation of a skyrmionium with $Q=0$ into a skyrmion with $Q=+1$ in the nanodisk. On the other hand, when the magnetic field is applied perpendicularly to the nanodisk plane and along the $-z$-direction, that is, $H_{z}<0$ mT, the skyrmionium size expands, where the radius of the inner circular domain wall decreases and that of the outer one increases. When the amplitude of the magnetic field is reduced to be smaller than a certain threshold, that is, $H_{z}=-260$ mT, the outer circular domain wall will be eliminated from the edge of the nanodisk, giving rise to the transformation of a skyrmionium with $Q=0$ into a skyrmion with $Q=-1$ in the nanodisk. It should be noted that if the amplitude of the applied magnetic field is larger than $450$ mT or smaller than $-490$ mT, that is, $H_{z}>450$ mT or $H_{z}<-490$ mT, the nanodisk will be magnetized to its saturated state, that is, the FM state.

\begin{figure}[t]
\centerline{\includegraphics[width=0.50\textwidth]{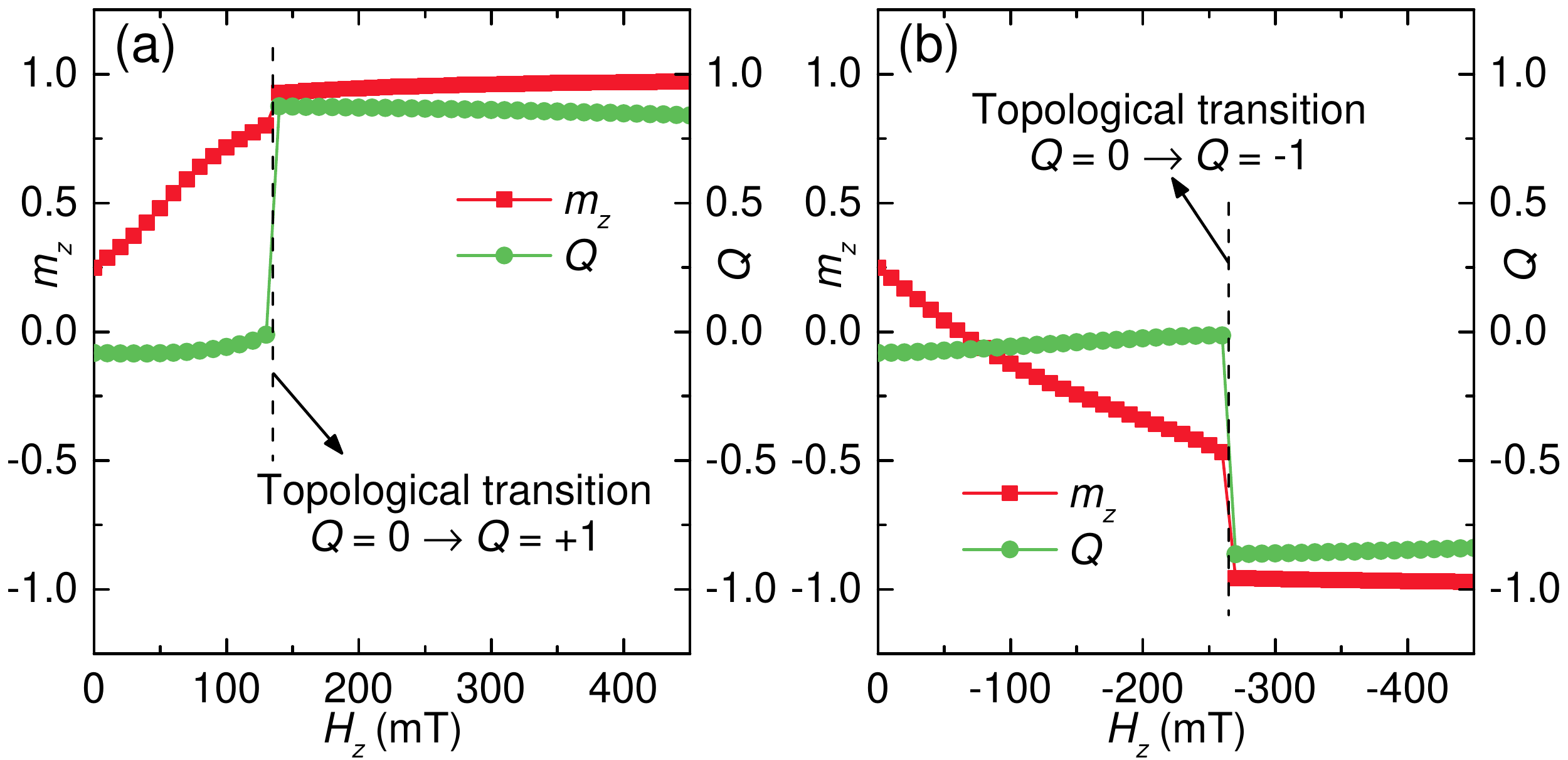}}
\caption{(Color online)
The normalized out-of-plane magnetization component $m_z$ and the topological number $Q$ as a function of $H_{z}$ applied along (a) the $+z$-direction and (b) the $-z$-direction, corresponding to Fig.~\ref{FIG5}(c).
}
\label{FIG6}
\end{figure}

We call the magnetic field-induced transformation of a skyrmionium with $Q=0$ into a skyrmion with either $Q=+1$ or $Q=-1$ as the degeneration of a skyrmionium. As shown in Fig.~\ref{FIG5}(c), we represent the profiles of the relaxed out-of-plane magnetization component along the nanodisk diameter [see Fig.~\ref{FIG5}(a)] at successively increasing or decreasing applied magnetic field, which are measured with a field step of $10$ mT in the range of $H_{z}=-490\sim 450$ mT. It clearly illustrates the irreversible degeneration of a skyrmionium with the topological number transition of $Q=0\rightarrow Q=\pm 1$. Figures~\ref{FIG6}(a) and \ref{FIG6}(b) show the normalized out-of-plane magnetization component $m_z$ and the topological number $Q$ of the nanodisk as functions of $H_z$ applied along the $+z$-direction and the $-z$-direction, respectively, corresponding to Fig.~\ref{FIG5}(c). It can be seen that $m_z$ is proportional to $H_z$ for a skyrmionium with $Q=0$, while it is almost constant for a skyrmion with $Q=\pm 1$. We also note there exist sharp changes of $m_z$ with respect to $H_z$ at the topological transitions of $Q=0\rightarrow Q=+1$ and $Q=0\rightarrow Q=-1$, as denoted in Fig.~\ref{FIG6}, which indicates the drastic change of magnetization configuration in the nanodisk at the critical moment of the skyrmionium degeneration.

It is worth mentioning that the degeneration of a skyrmionium in the nanodisk can also be triggered by other stimuli, such as the spin-polarized current. As presented in a recent study~\cite{Liu_PRB2015}, the skyrmionium is switched in the nanodisk by applying a spin-polarized current with the CPP geometry, which is a methodologically similar but topologically different process to the skyrmionium degeneration.

\section{Motion of a skyrmionium}
\label{se:Motion}

We have shown the spin-polarized current-induced generation and magnetic field-driven manipulation of a skyrmionium in Sec.~\ref{se:Generation} and Sec.~\ref{se:Manipulation}, respectively. Now we move on to discuss the spin-polarized current-driven motion of a skyrmionium, its distortion, as well as its destruction in a magnetic nanotrack.

We first simulate the spin-polarized current-driven motion of a skyrmionium in the nanotrack with $l=1000$ nm, $w=150$ nm, and $D=3.5$ mJ m$^{-2}$. A skyrmionium is created and relaxed near the left end of the nanotrack, that is, with its center at $(0.1l, 0.5w)$, before we applying the driving current. For the CPP geometry, a spin-polarized current is injected into the nanotrack from the bottom, which can be realized by the spin Hall effect~\cite{Sampaio_NNANO2013,Tomasello_SREP2014,Wanjun_SCIENCE2015}. For the CIP geometry, a spin-polarized current is injected into the nanotrack from the left end, that is, the electrons flow toward the right in the nanotrack. For contrast and comparison purposes, we also simulate the spin-polarized current-driven motion of a skyrmion in the same device.

\begin{figure}[t]
\centerline{\includegraphics[width=0.50\textwidth]{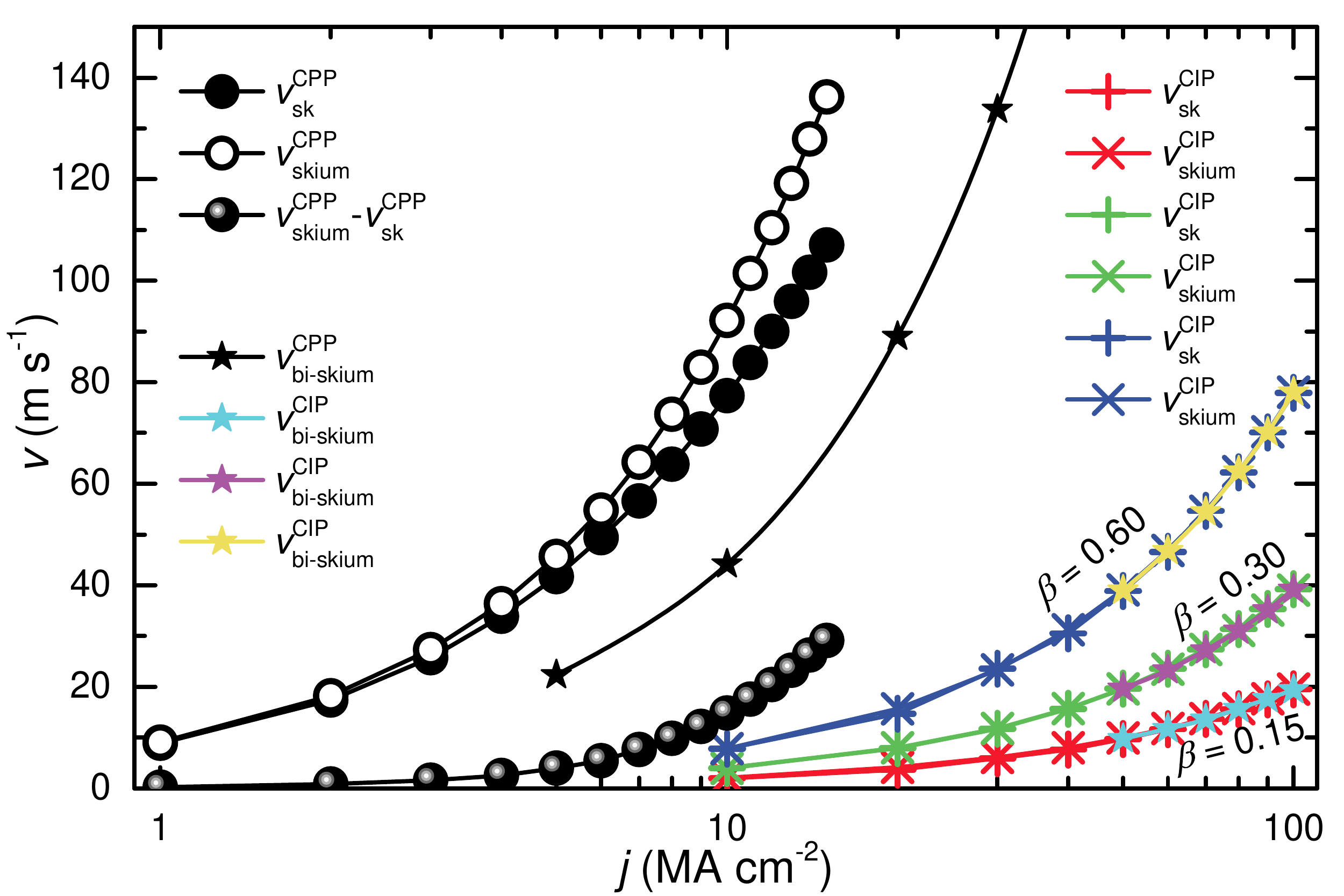}}
\caption{(Color online)
Velocity for the skyrmion $v_{\text{sk}}$, skyrmionium $v_{\text{skium}}$, and bilayer-skyrmionium $v_{\text{bi-skium}}$ driven by the spin-polarized current with the CPP or CIP geometry as a function of the driving current density $j$ in the nanotrack with $l=1000$ nm, $w=150$ nm, and $D=3.5$ mJ m$^{-2}$. The velocity at different $j$ is measured when steady motion is attained. The skyrmionium and skyrmion will be destroyed when the driving current density with the CPP geometry is larger than $15$ MA cm$^{-2}$ and $17$ MA cm$^{-2}$, respectively.
}
\label{FIG7}
\end{figure}

As shown in Fig.~\ref{FIG7}, the skyrmion velocity $v_{\text{sk}}$ and the skyrmionium velocity $v_{\text{skium}}$ as functions of $j$ are micromagnetically calculated, where different results are found for the model with the CIP geometry and that with the CPP geometry. For the case of the CIP geometry, the skyrmionium and skyrmion attain the same current-velocity ($j$-$v$) relation for the given range of driving current density, that is, $j=10\sim 100$ MA cm$^{-2}$, which shows up as overlapping $j$-$v$ curves for the skyrmionium and skyrmion. Indeed, both $v_{\text{skium}}$ and $v_{\text{sk}}$ at a given $j$ increase with the strength of non-adiabatic STT torque $\beta$. The skyrmionium moves along the nanotrack without obvious distortion and reaches a steady velocity of $v_{\text{skium}}=78$ m s$^{-1}$ at $j=100$ MA cm$^{-2}$ and $\beta=0.6$.

By contrast, for the case of the CPP geometry, $v_{\text{skium}}$ is clearly higher than $v_{\text{sk}}$ when $j>2$ MA cm$^{-2}$, as shown in Fig.~\ref{FIG7}. It can be seen that the velocity difference between $v_{\text{skium}}$ and $v_{\text{sk}}$ increases with increasing $j$. The skyrmionium moves along the nanotrack without obvious distortion and reaches a steady velocity of $v_{\text{skium}}=46$ m s$^{-1}$ at $j=5$ MA cm$^{-2}$, where the skyrmion reaches a relatively low steady velocity of $v_{\text{sk}}=42$ m s$^{-1}$. The velocity difference is $v_{\text{skium}}-v_{\text{sk}}=4$ m s$^{-1}$. At $j=15$ MA cm$^{-2}$, the skyrmionium moves with an obvious distortion and reaches a steady velocity of $v_{\text{skium}}=136$ m s$^{-1}$, where the skyrmion only reaches a steady velocity of $v_{\text{sk}}=107$ m s$^{-1}$. The velocity difference is increased to be $v_{\text{skium}}-v_{\text{sk}}=29$ m s$^{-1}$. When the driving current density with the CPP geometry $j>15$ MA cm$^{-2}$, the moving skyrmionium will be extremely distorted and destroyed, which is caused by the SkHE exerted on the skyrmion with $Q=+1$ and the skyrmion with $Q=-1$ consisting the skyrmionium (see Fig.~\ref{FIG1}). Similarly, when the driving current density with the CPP geometry $j>17$ MA cm$^{-2}$, the moving skyrmion will be destroyed at the upper edge of the nanotrack due to the SkHE.

\begin{figure}[t]
\centerline{\includegraphics[width=0.50\textwidth]{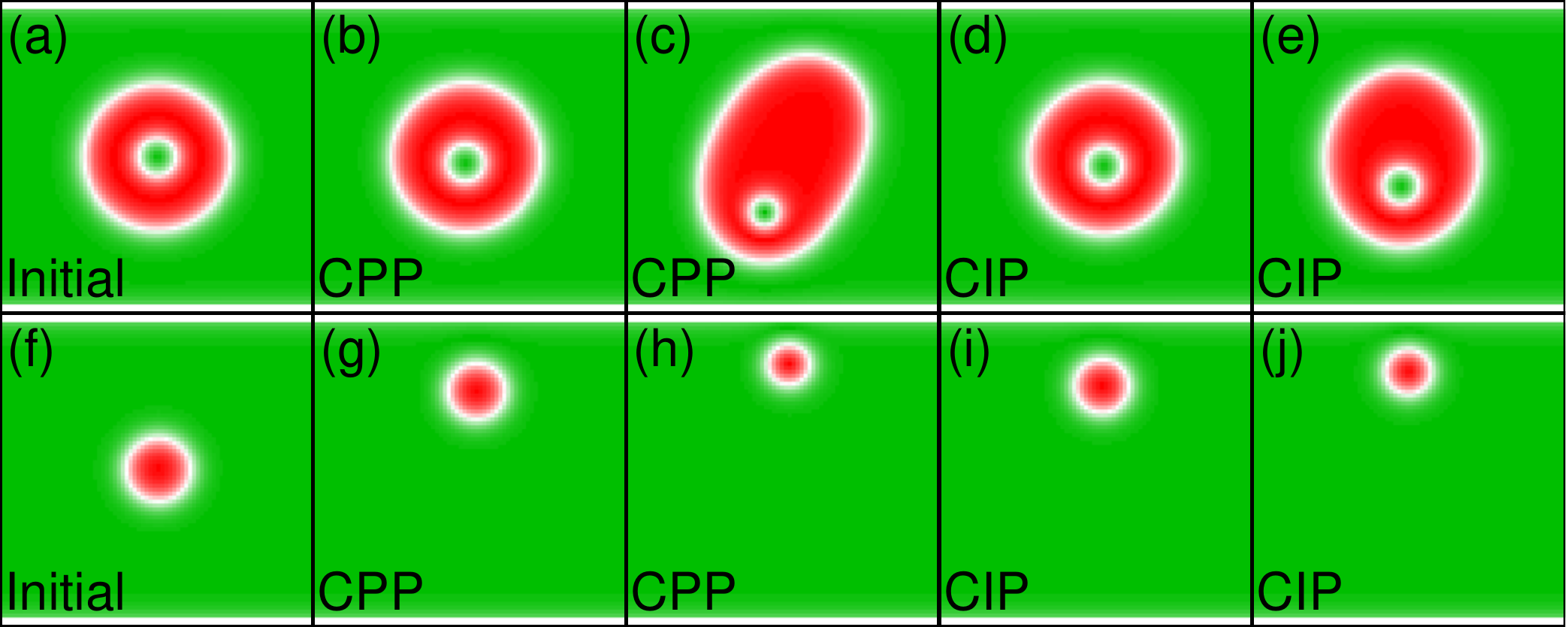}}
\caption{(Color online)
(a) A relaxed skyrmionium at $j=0$ MA cm$^{-2}$.
(b) A skyrmionium in the steady motion driven by the CPP geometry of $j=3$ MA cm$^{-2}$, and (c) of $j=13$ MA cm$^{-2}$.
(d) A skyrmionium in the steady motion driven by the CIP geometry of $j=90$ MA cm$^{-2}$, and (e) of $j=200$ MA cm$^{-2}$.
(f) A relaxed skyrmion at $j=0$ MA cm$^{-2}$.
(g) A skyrmion in the steady motion driven by the CPP geometry of $j=3$ MA cm$^{-2}$, and (h) of $j=13$ MA cm$^{-2}$.
(i) A skyrmion in the steady motion driven by the CIP geometry of $j=90$ MA cm$^{-2}$, and (j) of $j=200$ MA cm$^{-2}$. Here, for the CIP geometry, $\beta=0.6$.
}
\label{FIG8}
\end{figure}

The top-views of the skyrmionium and skyrmion driven by the CPP and the CIP geometries are illustrated in Fig.~\ref{FIG8}. Both the skyrmionium and skyrmion in the given nanotrack at zero driving current are relaxed at the central line of the nanotrack [see Figs.~\ref{FIG8}(a) and \ref{FIG8}(f)]. When a relatively small driving current of $j=3$ MA cm$^{-2}$ with the CPP geometry is applied, the skyrmionium moves along the central line of the nanotrack, while the skyrmion first approaches the upper edge and then moves along the edge. The top-views of their configurations in the steady motion are given in Figs.~\ref{FIG8}(b) and \ref{FIG8}(g), respectively. When a relatively large driving current of $j=13$ MA cm$^{-2}$ with the CPP geometry is applied, it can be seen that the skyrmionium still moves along the central line of the nanotrack, however, its structure is obviously distorted [see Fig.~\ref{FIG8}(c)]. The behavior of the skyrmion is the same as that driven by the relatively small driving current with the CPP geometry, although it presses more strongly the upper edge due to the presence of the relatively large driving current. On the other hand, when a relatively small driving current of $j=90$ MA cm$^{-2}$ with $\beta=0.6$ with the CIP geometry is applied, the skyrmionium moves along the central line of the nanotrack with no transverse shift [see Fig.~\ref{FIG8}(d)], and the skyrmion moves along the upper edge [see Fig.~\ref{FIG8}(i)]. When a relatively large driving current of $j=200$ MA cm$^{-2}$ with $\beta=0.6$ with the CIP geometry is applied, it shows that the skyrmionium moves along the nanotrack with an obvious distortion [see Fig.~\ref{FIG8}(e)], and the skyrmion moves along the upper edge with an increased transverse shift [see Fig.~\ref{FIG8}(j)].

By comparing the results of the CPP geometry and the CIP geometry, it is clear that the CPP geometry is significantly efficient in moving a skyrmionium in the nanotrack, which is the same as in the case of current-driven skyrmions~\cite{Sampaio_NNANO2013} and domain walls~\cite{Thiaville_EPL2012}. In order to understand the current-induced motion dynamics of the skyrmionium with the CPP and CIP geometries, we analyze the steady motion of the skyrmionium in the nanotrack using the Thiele equation approach~\cite{Thiele_PRL1973,Iwasaki_NC,Iwasaki_NL,Tomasello_SREP2014,He_PRB2006}, where we assume the skyrmionium is rigid in the steady motion.

\begin{figure*}[t]
\centerline{\includegraphics[width=1.00\textwidth]{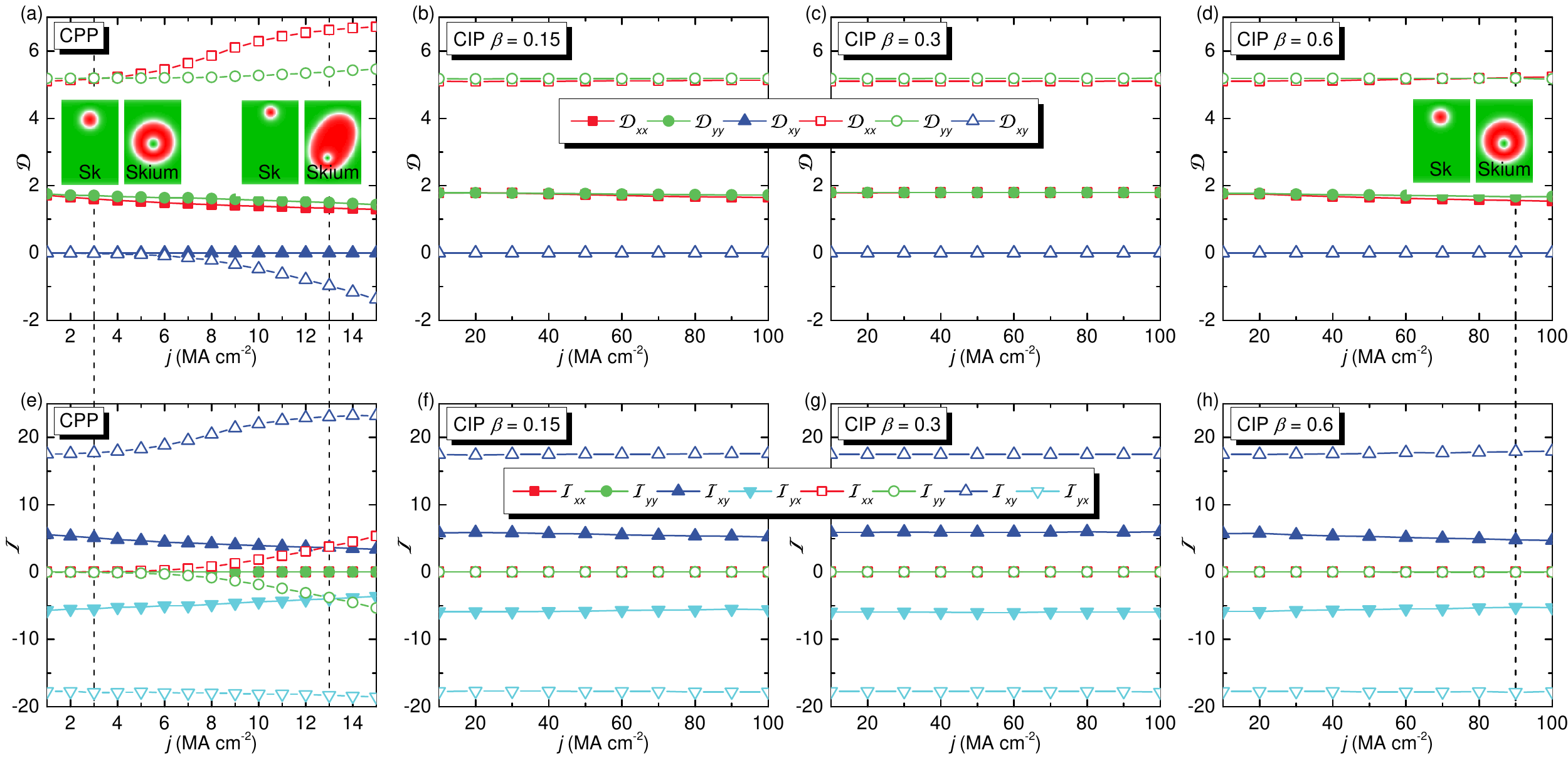}}
\caption{(Color online)
The components of the dissipative tensor $\boldsymbol\D$, that is, $\D_{xx}$, $\D_{yy}$, $\D_{xy}$, and $\D_{yx}$ as functions of $j$ for the skyrmion (solid symbols) and skyrmionium (open symbols) driven by (a) CPP geometry, (b) CIP geometry with $\beta=0.15$, (c) CIP geometry with $\beta=0.3$, and (d) CIP geometry with $\beta=0.6$. Here, $\D_{xy}=\D_{yx}$.
The components of the $\boldsymbol{\I}$ tensor, that is, $\I_{xx}$, $\I_{yy}$, $\I_{xy}$, and $\I_{yx}$ as functions of $j$ for the skyrmion (solid symbols) and skyrmionium (open symbols) driven by (e) CPP geometry, (f) CIP geometry with $\beta=0.15$, (g) CIP geometry with $\beta=0.3$, and (h) CIP geometry with $\beta=0.6$.
Insets show the skyrmion and skyrmionium at selected $j$ and current injection geometries indicated by the vertical dash lines.
}
\label{FIG9}
\end{figure*}

For the CPP geometry, the Thiele motion equation is derived from the magnetization dynamics equation (\ref{eq:LLGS-CPP}), which is expressed as follows,
\begin{equation}
\boldsymbol{G}\times\boldsymbol{v}^{\text{CPP}}-\alpha\boldsymbol{\D}\cdot\boldsymbol{v}^{\text{CPP}}+4\pi\boldsymbol{\B}\cdot\boldsymbol{j}+\boldsymbol{F}=\boldsymbol{0},
\label{eq:TME-CPP}
\end{equation}
where $\boldsymbol{v}^{\text{CPP}}=(v_x^{\text{CPP}},v_y^{\text{CPP}})$ is the steady velocity of the skyrmionium or skyrmion, and $\boldsymbol{j}=(j,0)$ is the driving current. $\boldsymbol{F}$ stands for the force exerted on the skyrmionium or skyrmion by the edge of the nanotrack. For the situation that the skyrmionium and skyrmion have reached the steady motion in the nanotrack, we have two relations hold directly: $v_{y}^\text{CPP}=0$ and $F_{x}=0$ (see Refs.~\onlinecite{Iwasaki_NNANO2013,Iwasaki_NL,Sampaio_NNANO2013}). The first term on left-hand side of Eq.~(\ref{eq:TME-CPP}) is the Magnus force term with the gyromagnetic coupling vector
\begin{equation}
\boldsymbol{G}=(0,0,4\pi Q),
\label{eq:G-vector}
\end{equation}
where $Q=0$ for the skyrmionium and $Q=\pm 1$ for the skyrmion. The second term on left-hand side of Eq.~(\ref{eq:TME-CPP}) is the dissipative force term with the dissipative tensor describing the effect of the dissipative force on the moving skyrmionium or skyrmion
\begin{equation}
\boldsymbol{\D}=4\pi\begin{pmatrix} \D_{xx} & \D_{xy} \\ \D_{yx} & \D_{yy} \end{pmatrix},
\label{eq:D-tensor}
\end{equation}
where the tensor components are calculated by
\begin{equation}
\D_{\mu\nu}=\frac{1}{4\pi}\int\frac{\partial\boldsymbol{m}}{\partial\mu}\cdot\frac{\partial\boldsymbol{m}}{\partial\nu}dxdy,
\label{eq:Dxx-Dyy-Dxy-Dyx}
\end{equation}
where $\mu$ and $\nu$ run over $x$ and $y$. The integrations are carried out over the region containing the skyrmionium or the skyrmion. The third term on left-hand side of Eq.~(\ref{eq:TME-CPP}) is the driving force term with the tensor linked to the STT (see Sec.~\ref{se:Methods})
\begin{equation}
\boldsymbol{\B}=\frac{u}{aj}\begin{pmatrix} -\I_{xy} & \I_{xx} \\ -\I_{yy} & \I_{yx} \end{pmatrix},
\label{eq:B-tensor}
\end{equation}
where the tensor components are calculated by
\begin{equation}
\I_{\mu\nu}=\frac{1}{4\pi}\int\left(\frac{\partial\boldsymbol{m}}{\partial\mu}\times\boldsymbol{m}\right)_{\nu}dxdy,
\label{eq:Ixx-Iyy-Ixy-Iyx}
\end{equation}
where $\mu$ and $\nu$ run over $x$ and $y$. 
Similarly, the integrations are carried out over the region containing the skyrmionium or the skyrmion. For the skyrmionium in the given nanotrack [see Fig.~\ref{FIG8}(a)], we find it has $\D_{xx}=\D_{yy}=5$, $\D_{xy}=\D_{yx}=0$, $\I_{xx}=\I_{yy}=0$ and $\I_{xy}=-\I_{yx}=36a$. For the skyrmion in the given nanotrack [see Fig.~\ref{FIG8}(f)], we find it has $\D_{xx}=\D_{yy}=2$, $\D_{xy}=\D_{yx}=0$, $\I_{xx}=\I_{yy}=0$ and $\I_{xy}=-\I_{yx}=12a$. Hence, from the Thiele motion equation (\ref{eq:TME-CPP}) for the CPP geometry we find
\begin{equation}
v_x^{\text{CPP}}= \frac{u \I_{xy}}{a\alpha \D_{xx}}, \quad
v_y^{\text{CPP}}= 0.
\label{eq:TME-CPP-vx-vy}
\end{equation}

By substituting the parameters of the skyrmionium with $Q=0$ into Eq.~(\ref{eq:TME-CPP-vx-vy}), we find the steady velocity of the skyrmionium as
\begin{equation}
v_{x,\text{skium}}^{\text{CPP}}=\frac{7.2u}{\alpha}, \quad
v_{y,\text{skium}}^{\text{CPP}}= 0.
\label{eq:skium-CPP-vx-vy}
\end{equation}
In the same way, by substituting the parameters of the skyrmion with $Q=+1$ into Eq.~(\ref{eq:TME-CPP-vx-vy}), we find the steady velocity of the skyrmion as
\begin{equation}
v_{x,\text{sk}}^{\text{CPP}}=\frac{6u}{\alpha}, \quad
v_{y,\text{sk}}^{\text{CPP}}=0.
\label{eq:sk-CPP-vx-vy}
\end{equation}
the skyrmionium and skyrmion go straight in the $x$-direction as they have $v_{y}^{\text{CPP}}=0$ in the steady motion. Thus the velocity difference between the skyrmionium and skyrmion at a certain $j$ for our parameters, where $\alpha=0.3$, is given as
\begin{equation}
v_{\text{skium}}^{\text{CPP}}-v_{\text{sk}}^{\text{CPP}}=4u.
\label{eq:skium-sk-CPP-v-diff}
\end{equation}
Clearly, it can be seen that, with the CPP geometry, $v_{\text{skium}}^{\text{CPP}}$ is larger than $v_{\text{sk}}^{\text{CPP}}$ when $j>0$ MA cm$^{-2}$, and the velocity difference is proportional to $j$, which matches well with our simulation results (see Fig.~\ref{FIG7}), although the skyrmionium in the simulation will be distorted at a large $j$ [see Fig.~\ref{FIG8}(c)]. It should be noted that, in the given nanotrack, the skyrmion approaches the edge as it has $v_{y,\text{sk}}^{\text{CPP}}\neq 0$ due to the SkHE at the beginning of the simulation. However, the edge of the nanotrack will limit its motion in the $y$-direction, and thus it will have a constant transverse shift [compare Fig.~\ref{FIG8}(f) and Fig.~\ref{FIG8}(g)] and $v_{y,\text{sk}}^{\text{CPP}}=0$ when the steady motion is reached.

For the CIP geometry, the Thiele motion equation is derived from the magnetization dynamics equation (\ref{eq:LLGS-CIP}), which is expressed as follows
\begin{equation}
\boldsymbol{G}\times\left(\boldsymbol{v}^\text{CIP}-\boldsymbol{u}\right)+\boldsymbol{\D}\left(\beta\boldsymbol{u}-\alpha\boldsymbol{v}^\text{CIP}\right)+\boldsymbol{F}=\boldsymbol{0},
\label{eq:TME-CIP}
\end{equation}
where $\boldsymbol{v}^{\text{CIP}}=(v_x^{\text{CIP}},v_y^{\text{CIP}})$ is the steady velocity of the skyrmionium or skyrmion, and $\boldsymbol{u}=(u,0)$ is determined by the driving current $j$ (see Sec.~\ref{se:Methods}). Similarly, for both the skyrmionium and skyrmion in the steady motion, we have two relations hold directly: $v_{y}^\text{CIP}=0$ and $F_{x}=0$ (see Refs.~\onlinecite{Iwasaki_NNANO2013,Iwasaki_NL,Sampaio_NNANO2013}). Hence, from Eq.~(\ref{eq:TME-CIP}) we find
\begin{equation}
v_{x}^\text{CIP}=\frac{\beta u}{\alpha}, \quad
v_{y}^\text{CIP}=0.
\label{eq:TME-CIP-vx-vy}
\end{equation}
It can be seen that, for both the skyrmionium and skyrmion driven by the CIP geometry in the given nanotrack, we have
\begin{equation}
v_{x,\text{skium}}^{\text{CIP}}=v_{x,\text{sk}}^{\text{CIP}}=\frac{\beta u}{\alpha}, \quad
v_{y,\text{skium}}^{\text{CIP}}=v_{y,\text{sk}}^{\text{CIP}}= 0.
\label{eq:skium-sk-CIP-vx-vy}
\end{equation}
the skyrmionium and skyrmion in the given nanotrack undergo a straight steady motion in the $x$-direction as they have $v_{y}^{\text{CIP}}=0$. The velocity difference between the skyrmionium and skyrmion at a certain $j$ is straightforwardly given as
\begin{equation}
v_{\text{skium}}^{\text{CIP}}-v_{\text{sk}}^{\text{CIP}}=0.
\label{eq:skium-sk-CIP-v-diff}
\end{equation}
That is to say, in the presence of the driving current with the CIP geometry, the steady velocity of the skyrmionium is identical to that of the skyrmion at a certain $j$, which is in good agreement with our simulation results (see Fig.~\ref{FIG7}).

On the other hand, by comparing Eq.~(\ref{eq:skium-CPP-vx-vy}) and Eq.~(\ref{eq:skium-sk-CIP-vx-vy}), we calculate the steady-velocity ratio of the skyrmionium driven by the CPP geometry to that driven by the CIP geometry, which is expressed as
\begin{equation}
v_{\text{skium}}^{\text{CPP}}/v_{\text{skium}}^{\text{CIP}}=7.2/\beta.
\label{eq:skium-CPP-CIP-v-ratio}
\end{equation}
It shows that the ratio only depends on the strength of non-adiabatic STT torque $\beta$. For example, the steady velocity of the skyrmionium driven by the CPP geometry is $24$ times higher than that driven by the CIP geometry at a certain $j$ when $\beta=0.3$. In our simulation with $\beta=0.3$ and $j=10$ MA cm$^{-2}$ (see Fig.~\ref{FIG7}), we numerically find $v_{\text{skium}}^{\text{CPP}}=92$ m s$^{-1}$ and $v_{\text{skium}}^{\text{CIP}}=3.9$ m s$^{-1}$, where we have $v_{\text{skium}}^{\text{CPP}}/v_{\text{skium}}^{\text{CIP}}=23.6$, which matches well with the analytical solution $v_{\text{skium}}^{\text{CPP}}/v_{\text{skium}}^{\text{CIP}}=24$ given by Eq.~(\ref{eq:skium-CPP-CIP-v-ratio}).

\begin{figure}[t]
\centerline{\includegraphics[width=0.50\textwidth]{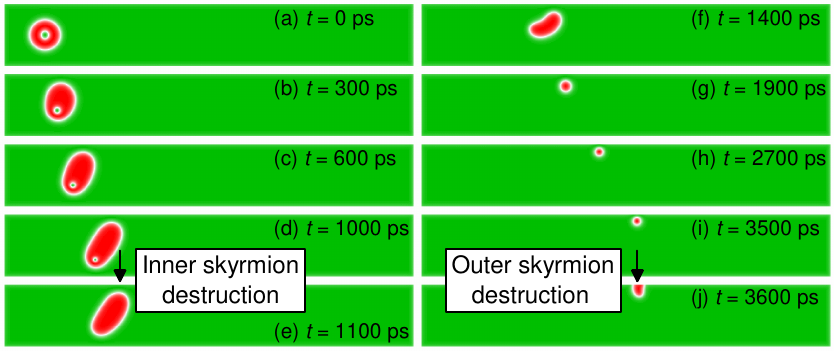}}
\caption{(Color online)
The destruction process of a skyrmionium driven by the CPP geometry of $j=18$ MA cm$^{-2}$ in the nanotrack with $l=1000$ nm, $w=150$ nm, and $D=3.5$ mJ m$^{-2}$. The skyrmionium goes straight with deformation since the opposite Magnus forces act on the outer skyrmion with $Q=+1$ and the inner skyrmion with $Q=-1$, which results in the destruction of the skyrmionium into a skyrmion with $Q=+1$. After the destruction, the skyrmion with $Q=+1$ touches the upper edge of the nanotrack as the SkHE is active.
}
\label{FIG10}
\end{figure}

\begin{figure}[t]
\centerline{\includegraphics[width=0.50\textwidth]{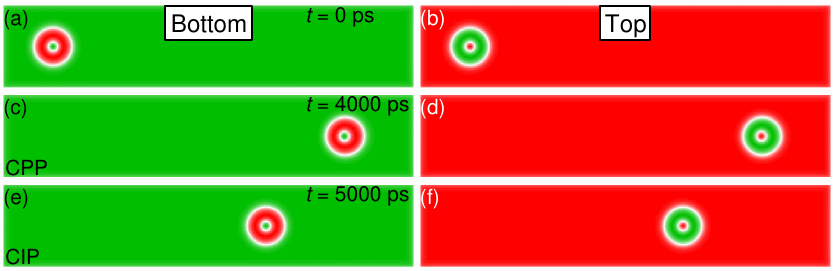}}
\caption{(Color online)
Top-views of the motion of a magnetic bilayer-skyrmionium driven by the CPP or CIP geometry in an antiferromagnetically exchange-coupled bilayer nanotrack with $l=750$ nm, $w=150$ nm, and $D=3.5$ mJ m$^{-2}$. Top-views of (a) the bottom and (b) the top FM layers at $t=0$ ps, where no driving current is applied. Top-views of (c) the bottom and (d) the top FM layers when the CPP driving current of $j=30$ MA cm$^{-2}$ is applied, the bilayer-skyrmionium moves along the nanotrack without any distortion, reaching a steady velocity of $v_{\text{bi-skium}}=134$ m s$^{-1}$. Top-views of (e) the bottom and (f) the top FM layers when the driving current of $j=100$ MA cm$^{-2}$ with the CIP geometry is applied with $\beta=0.6$, the bilayer-skyrmionium moves along the nanotrack without any distortion, reaching a steady velocity of $v_{\text{bi-skium}}=78$ m s$^{-1}$.
}
\label{FIG11}
\end{figure}

Although we assume the skyrmionium is a rigid object in the analysis using the Thiele motion equations, it should be noted that the skyrmionium will be distorted and even be destroyed when $j$ is larger than a certain value, either with the CPP geometry or CIP geometry [see Figs.~\ref{FIG8}(c) and \ref{FIG8}(e)]. The reason is that the skyrmion with $Q=+1$ and the skyrmion with $Q=-1$, which consist the skyrmionium, experience opposite forces induced by the driving current, which are proportional to $j$. Here we can find the edge-induced force acted on the skyrmion in the steady motion driven by the CPP geometry from Eq.~(\ref{eq:TME-CPP}),
\begin{equation}
F_{x}^{\text{CPP}}=0, \quad
F_{y}^{\text{CPP}}=-\frac{4\pi u Q \I_{xy}}{a\alpha \D_{xx}},
\label{eq:TME-CPP-Fx-Fy}
\end{equation}
as well as the edge-induced force acted on the skyrmion in the steady motion driven by the CIP geometry from Eq.~(\ref{eq:TME-CIP}),
\begin{equation}
F_{x}^{\text{CIP}}=0, \quad
F_{y}^{\text{CIP}}=\frac{4\pi u Q (\alpha-\beta)}{\alpha}.
\label{eq:TME-CIP-Fx-Fy}
\end{equation}
It is noteworthy that, with our simulation parameters, for example $\beta=0.6$, the ratio of the edge-induced force acted on a skyrmion with $Q=+1$ driven by the CPP geometry to that driven by the CIP geometry are given as
\begin{equation}
F_{y, \text{sk}}^{\text{CPP}}/F_{y, \text{sk}}^{\text{CIP}}=-\frac{\I_{xy}}{a \D_{xx} (\alpha-\beta)}=20.
\label{eq:sk-CPP-CIP-Fy-ratio}
\end{equation}
That is to say, the current-induced force under the CPP geometry, which is the reaction force of the edge-induced one, acted on the skyrmion at a certain $j$ and $\beta=0.6$ is $20$ times larger than that induced by the CIP geometry. This is the reason that the distortion of the skyrmionium [compare Fig.~\ref{FIG8}(c) and Fig.~\ref{FIG8}(d)] as well as the transverse shift of the skyrmion [compare Fig.~\ref{FIG8}(h) and Fig.~\ref{FIG8}(i)] driven by the CIP geometry are much smaller than that driven by the CPP geometry at a certain $j$.

\begin{figure}[t]
\centerline{\includegraphics[width=0.50\textwidth]{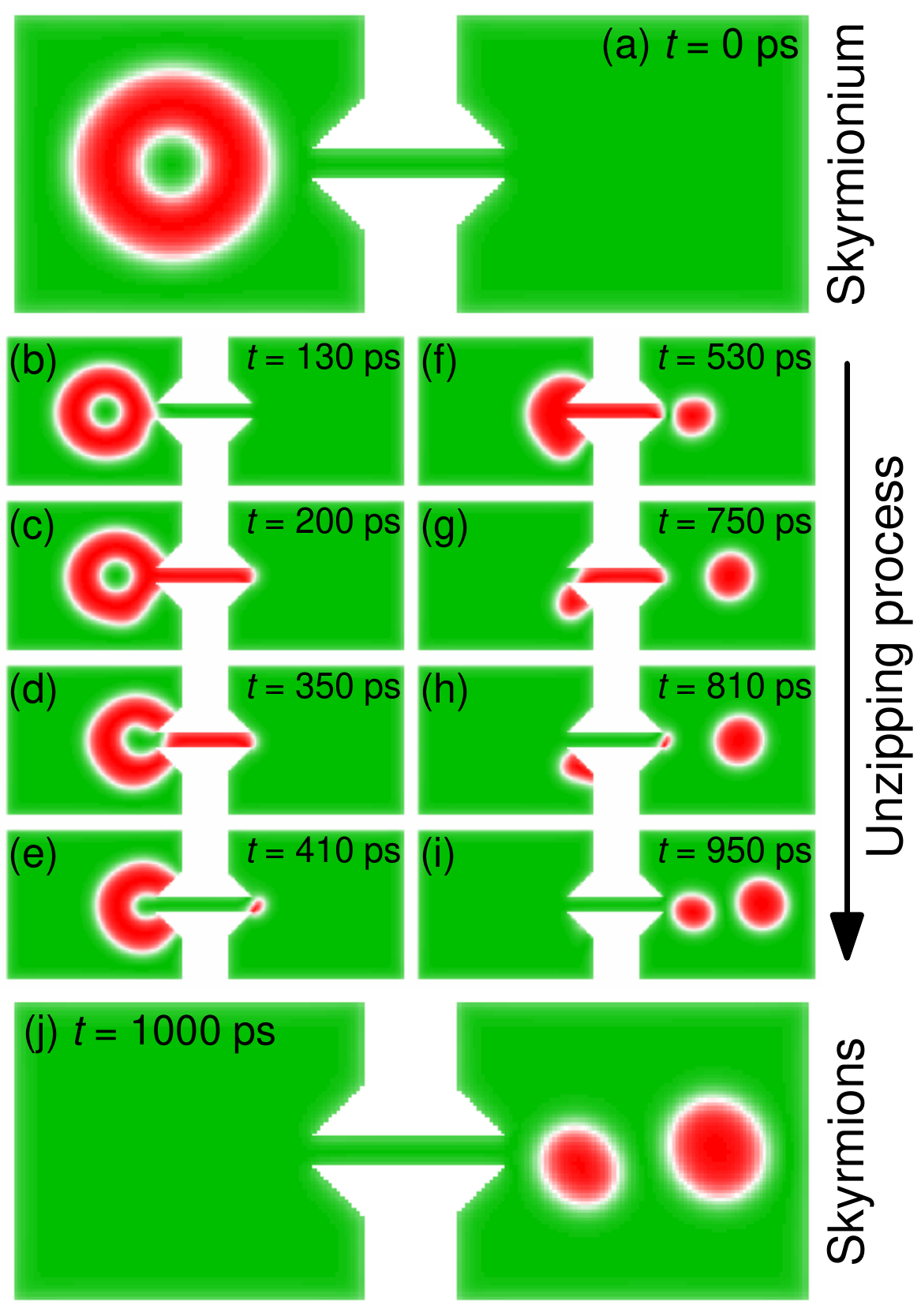}}
\caption{(Color online)
Unzipping of a skyrmionium with $Q=0$ into two skyrmions with $Q=+1$ in a nanotrack device with the junction geometry. (a)-(j) Top-views of the magnetization configuration of the device at selected $t$.
}
\label{FIG12}
\end{figure}

\begin{figure*}[t]
\centerline{\includegraphics[width=1.00\textwidth]{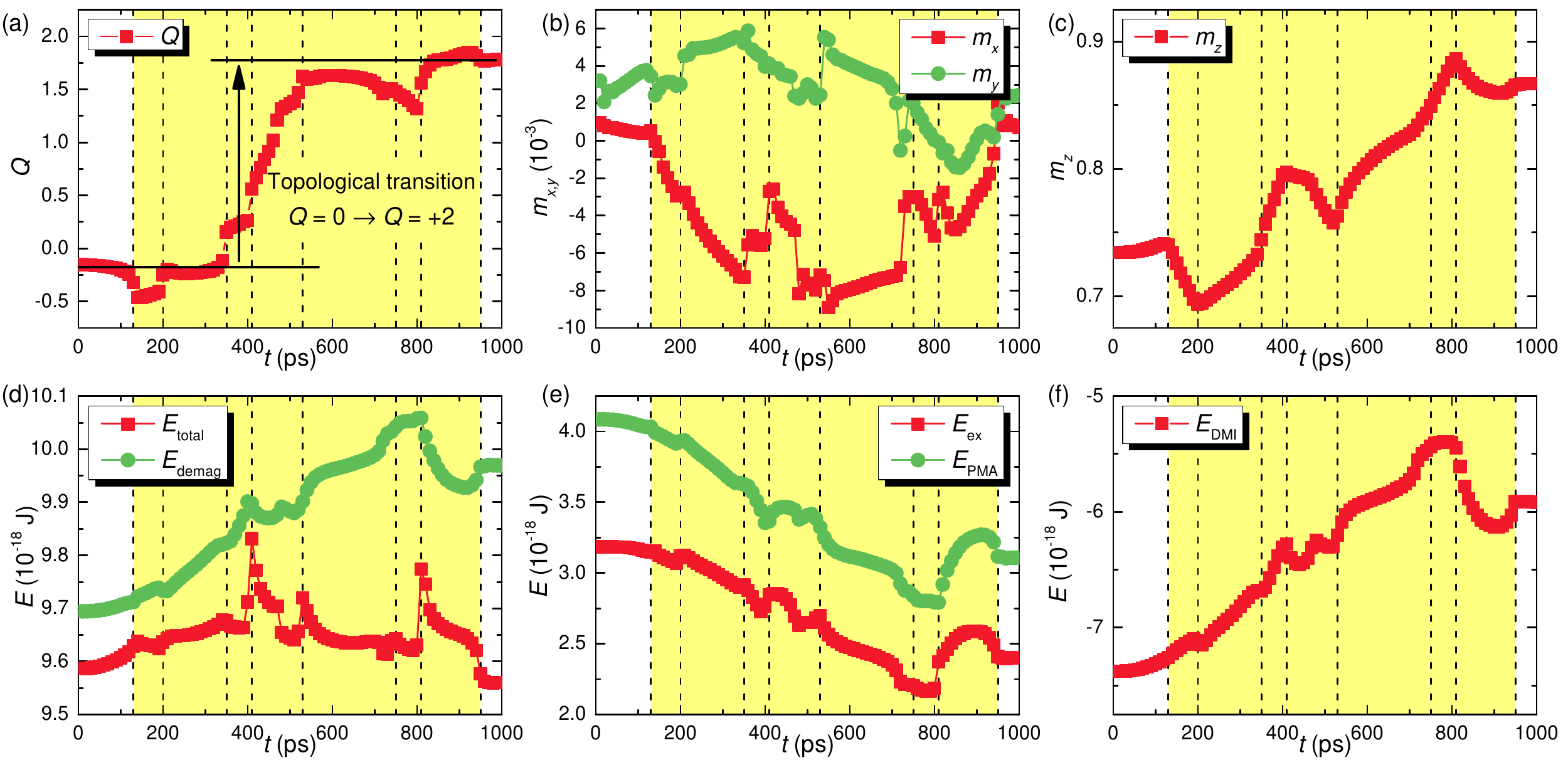}}
\caption{(Color online)
(a) The topological number $Q$, (b) the normalized in-plane magnetization components $(m_x,m_y)$, (c) the normalized out-of-plane magnetization component $m_z$, (d) the total micromagnetic energy $E_{\text{total}}$ and demagnetization energy $E_{\text{demag}}$, (e) the exchange energy $E_{\text{ex}}$ and PMA energy $E_{\text{PMA}}$, and (f) the DMI energy $E_{\text{DMI}}$ as functions of $t$. The topological transition of $Q=0\rightarrow Q=+2$ occurs within $1000$ ps, indicating the success of the unzipping event. Vertical dashed lines indicate the selected times corresponding to the snapshots shown in Fig.~\ref{FIG12}. Light yellow background indicates the time span of the unzipping process.
}
\label{FIG13}
\end{figure*}

Here, in order to further analyze a deformed skyrmionium, we calculated the $\boldsymbol{\D}$ tensor components as well as the $\boldsymbol{\I}$ tensor components as functions of time for different current injection geometries. For the CPP geometry, the components of the $\boldsymbol{\D}$ and $\boldsymbol{\I}$ tenors for the skyrmionium are found to vary with time when $j>6$ MA cm$^{-2}$, indicating the distortion of the skyrmionium and the deviation of the Thiele equations solutions from the simulation results [see Figs.~\ref{FIG9}(a) and \ref{FIG9}(e)]. Obviously, the components of the $\boldsymbol{\D}$ and $\boldsymbol{\I}$ tenors for the skyrmion vary little with time, indicating its distortion is much smaller as compared to that of the skyrmionium at a certain $j$. For the CIP geometry, as shown in Figs.~\ref{FIG9}(b-d) and \ref{FIG9}(f-h), the components of the $\boldsymbol{\D}$ and $\boldsymbol{\I}$ tenors for both the skyrmionium and skyrmion are almost constant with time when $j<100$ MA cm$^{-2}$, which justifies that the distortion of the skyrmionium driven by the CIP geometry is much smaller than that driven by the CPP geometry at a certain $j$.

Figure~\ref{FIG10} demonstrates the destruction process of a skyrmionium driven by the CPP geometry of $j=18$ MA cm$^{-2}$. It shows that the skyrmionium is obviously distorted once the driving current is applied. Due to the opposite forces exerted on the inner and outer skyrmions of the skyrmionium, the inner skyrmion with $Q=-1$ shrinks and the outer skyrmion with $Q=+1$ expanses, leading to the destruction of the inner skyrmion with $Q=-1$ [see Figs.~\ref{FIG10}(d) and \ref{FIG10}(e)]. The destruction is possible because of the lattice structure inherent to the system~\cite{Xichao_PRB2016}. Hence the skyrmionium with $Q=0$ is transformed into a skyrmion with $Q=+1$, which continues to move rightward, and then is destroyed by touching the upper edge since the SkHE is active [see Figs.~\ref{FIG10}(i) and \ref{FIG10}(j)].

\section{Motion of a bilayer-skyrmionium}
\label{se:Bilayer-Skyrmionium}

Because the distortion and destruction of the skyrmionium in the high-speed operation are detrimental to practical applications, we therefore construct the bilayer-skyrmionium in a nanotrack consisting of two antiferromagnetically exchange-coupled FM layers, which is supposed to have no distortion in motion, similar to the bilayer-skyrmion reported in Ref.~\onlinecite{Xichao_NCOMMS2016}. The nanotrack, with $l=750$ nm and $w=150$ nm, consists of three layers, that is, the bottom and top FM layers with $D=3.5$ mJ m$^{-2}$ and the middle nonmagnetic spacer. The thickness of each layer equals $a$. We assume the interlayer antiferromagnetic exchange coupling constant is equal to $-A/15$. A bilayer-skyrmionium is first created and relaxed near the left end of the nanotrack, that is, with its center at $(0.12l, 0.5w)$, before we applying the driving current. For the CPP geometry, a spin-polarized current is injected into the bottom FM layer from the bottom. For the CIP geometry, a spin-polarized current is injected into both the bottom and the top FM layers from the left. Details on the CPP and CIP geometries used here are described in Ref.~\onlinecite{Xichao_NCOMMS2016}.

The steady velocity of the bilayer-skyrmionium $v_{\text{bi-skium}}$ as a function of $j$ for different current injection geometries is also given in Fig.~\ref{FIG7}. It can be seen that the bilayer-skyrmionium has the same $j-v$ relation of the skyrmionium when it is driven by the CIP geometry. However, $v_{\text{bi-skium}}$ induced by the CPP geometry is basically a half of $v_{\text{skium}}$ induced by the CPP geometry at a certain $j$. The reason is that the CPP geometry is only injected into the bottom FM layer, but drives both skyrmioniums in the bottom and the top FM layers. Similar results hold for the bilayer-skyrmion as investigated in Ref.~\onlinecite{Xichao_ARXIV2016}.

Figures~\ref{FIG11}(a) and \ref{FIG11}(b) show, respectively, the initial states of the bottom and the top FM layers of the bilayer nanotrack, where a relaxed bilayer-skyrmionium is located near the left end. Figures~\ref{FIG11}(c) and \ref{FIG11}(d) show the motion of the bilayer-skyrmionium driven by the CPP geometry of $j=30$ MA cm$^{-2}$, where the bilayer-skyrmionium moves along the nanotrack without any distortion, and reaches a steady velocity of $v_{\text{bi-skium}}=134$ m s$^{-1}$. Figures~\ref{FIG11}(e) and \ref{FIG11}(f) show the motion of the bilayer-skyrmionium driven by the CIP geometry of $j=100$ MA cm$^{-2}$ with $\beta=0.6$. The bilayer-skyrmionium also moves along the nanotrack without any distortion, reaching a steady velocity of $v_{\text{bi-skium}}=78$ m s$^{-1}$. It can be seen that the bilayer-skyrmionium driven by the CPP geometry can reach a steady high velocity (higher than $100$ m s$^{-1}$), and it will not be distorted in the high-speed motion. The reason is that the bottom and the top skyrmioniums are strongly coupled due to the interlayer antiferromagnetic exchange coupling, where the SkHEs acting on the skyrmions with $Q=+1$ and the skyrmions with $Q=-1$ consisting the bilayer-skyrmionium are totally suppressed.

\section{Unzipping of a skyrmionium}
\label{se:Unzipping}

In this section, motivated by the recent theoretical proposal of creating skyrmions by magnetic domain walls (see Refs.~\onlinecite{Yan_NCOMMS2014,Xichao_SREP2015B}) and its subsequent experimental realization (see Ref.~\onlinecite{Wanjun_SCIENCE2015}), we investigate the transformation process from a skyrmionium with $Q=0$ into one pair of skyrmions with $Q=+1$ in a nanotrack device, as demonstrated in Fig.~\ref{FIG12}. The device, with $D=3.7$ mJ m$^{-2}$, has a length of $l=400$ nm. It includes two wide nanotracks with $w=150$ nm, and a narrow nanotrack with $w=14$ nm. The thickness of the device equals $a$. By exploiting the junction geometry, one pair of skyrmions can be generated from a skyrmionium. A skyrmionium is first created and relaxed at the center of the left wide nanotrack, then the driving current of $j=400$ MA cm$^{-2}$ with $\beta=0.3$ with the CIP geometry is injected to the device from the left, driving the skyrmionium into motion toward the narrow nanotrack. The $j$ inside the narrow nanotrack is proportional to the $j$ inside the wide nanotrack with respect to the ratio of the wide width to the narrow width. No external magnetic field is applied in the process.

As shown in Fig.~\ref{FIG12}(a), the skyrmionium is relaxed at the center of the left wide nanotrack at $t=0$ ps. When the driving current with the CIP geometry is applied, the skyrmionium shifts toward the right direction. When the skyrmionium arrives at the junction interface at $t=130$ ps [see Fig.~\ref{FIG12}(b)], the outer peripheral spins are in touch with the sample boundary forming a magnetic domain wall in the narrow nanotrack, whereas the other part of the skyrmionium continues to move due to the driving force in the right wide nanotrack. The magnetic domain wall in the narrow nanotrack reaches the right terminal at $t=200$ ps [see Fig.~\ref{FIG12}(c)]. Then, the structure is deformed into a pair of two downward magnetic domains separated by upward background spins at $t=350$ ps [see Fig.~\ref{FIG12}(d)]. The two magnetic domain walls in the narrow nanotrack form a skyrmion at $t=410$ ps [see Fig.~\ref{FIG12}(e)]. Hence, the first created skyrmion enter the left wide nanotrack at $t=530$ ps [see Fig.~\ref{FIG12}(f)]. At $t=750$ ps, two new magnetic domain walls are formed in the narrow nanotrack [see Fig.~\ref{FIG12}(g)]. Therefore, the second skyrmion starts to be created at $t=810$ ps [see Fig.~\ref{FIG12}(h)]. The second created skyrmion enters the left wide nanotrack at $t=950$ ps [see Fig.~\ref{FIG12}(i)]. Thus a pair of skyrmions forms at $t=1000$ ps [see Fig.~\ref{FIG12}(j)], which can continue to propagate in the left wide nanotrack. We call this process as the unzipping of a skyrmionium.

The corresponding topological number of the device increases from $Q=0$ for the nontopological skyrmionium to $Q=+2$ for a pair of skyrmions, as illustrated in Fig.~\ref{FIG13}(a). The normalized in-plane magnetization components $(m_x,m_y)$ of the device oscillate within a very small amplitude during the unzipping process [see Fig.~\ref{FIG13}(b)], while the normalized out-of-plane magnetization component $m_{z}$ is increased when a pair of skyrmions is created [see Fig.~\ref{FIG13}(c)]. It can also be seen that the total micromagnetic energy of the device shows several sharp changes during the unzipping process [see Fig.~\ref{FIG13}(d)], which correspond to the critical moments where the skyrmionium is destructed and/or the skyrmions are formed as shown in Fig.~\ref{FIG12}. It should be noted that the vertical dashed lines in Fig.~\ref{FIG13} denote those critical moments shown in Fig.~\ref{FIG12}. The total micromagnetic energy remains basically the same value when the unzipping process has been completed, indicating that a skyrmionium and a skyrmion in the given device are on the same level of metastability. On the other hand, the demagnetization energy of the device increases during the unzipping process [see Fig.~\ref{FIG13}(d)], which may as a result of the dipole-dipole interaction between the two skyrmions. The exchange and PMA energies of the device decrease at the same time [see Fig.~\ref{FIG13}(e)], which indicates that the total length of the magnetic domain walls forming the two skyrmions are shorter than that forming the skyrmionium. For the same reason, the DMI energy of the device also increases during the unzipping process which shows the same profile as that of the demagnetization energy [see Fig.~\ref{FIG13}(f)].

\begin{figure}[t]
\centerline{\includegraphics[width=0.50\textwidth]{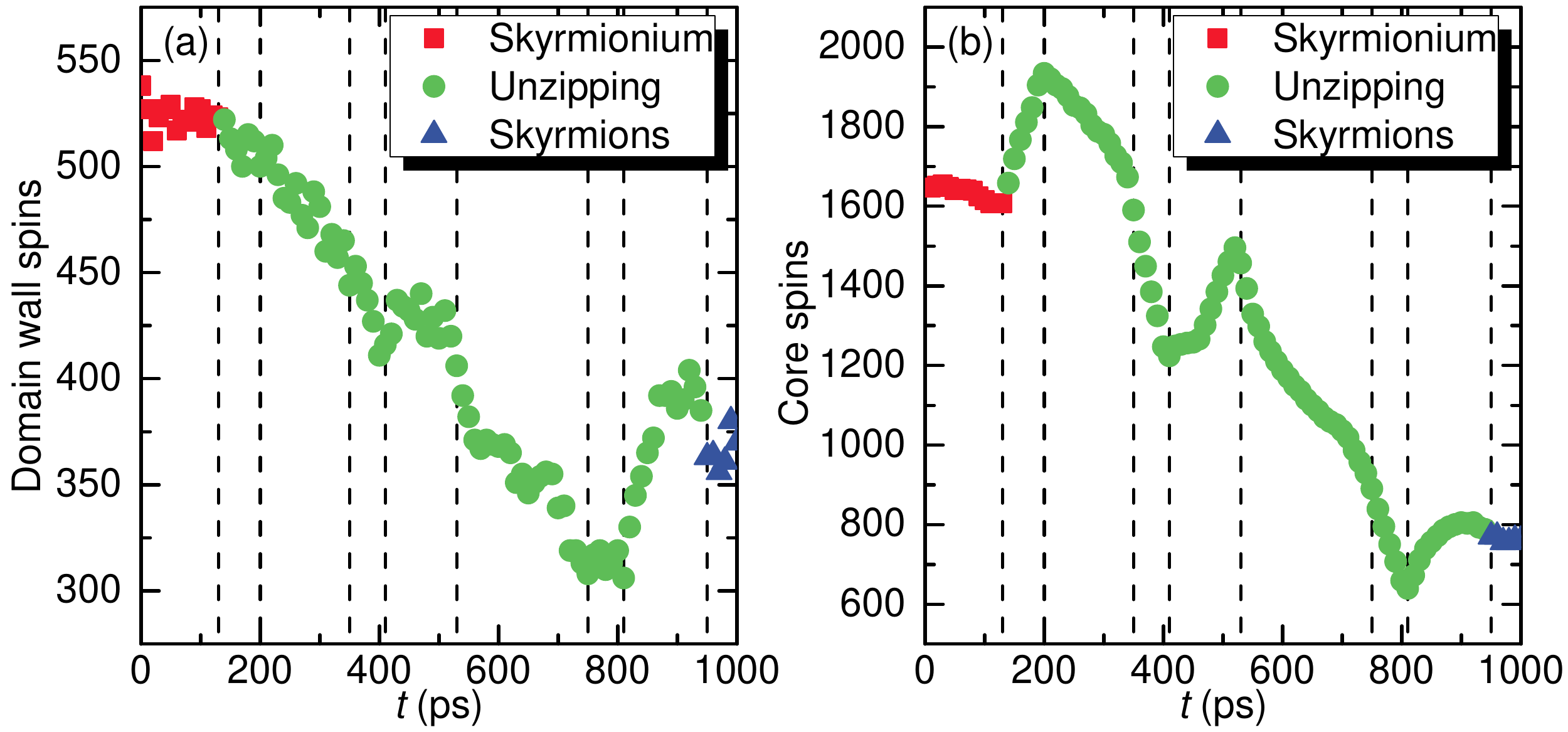}}
\caption{(Color online)
(a) Domain wall spins and (b) core spins as functions of $t$. The domain wall spins are defined as the spins which have $-0.5<m_{z}<0.5$, while the core spins are defined as the spins which have $m_{z}<0$. Square, circle and triangle symbols denote the data for the skyrmionium, unzipping process and skyrmions, respectively.
}
\label{FIG14}
\end{figure}

\begin{figure}[t]
\centerline{\includegraphics[width=0.50\textwidth]{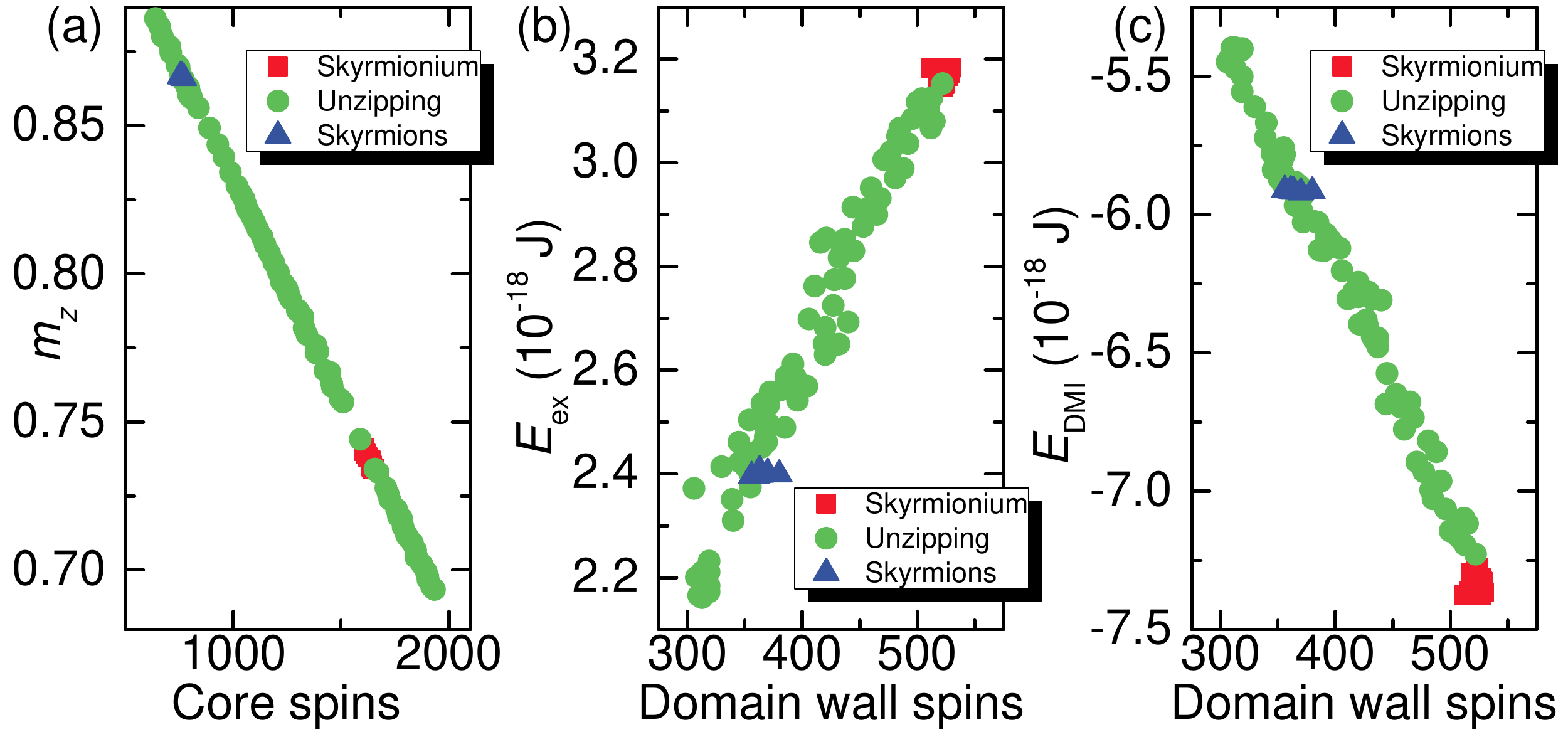}}
\caption{(Color online)
(a) $m_{z}$ as a function of core spins. (b) $E_{\text{ex}}$ and (c) $E_{\text{DMI}}$ as functions of domain wall spins. Square, circle and triangle symbols denote the data for the skyrmionium, unzipping process and skyrmions, respectively.
}
\label{FIG15}
\end{figure}

For the purpose of revealing the relations among the domain walls length, the energies, and the magnetization, we first calculate the number of domain wall spins as a function of time during the unzipping process [see Fig.~\ref{FIG14}(a)]. The number of the domain wall spins is proportional to the total length of the domain walls in the device, where the skyrmionium and/or skyrmions could be complete or broken (see Fig.~\ref{FIG12}). Similarly, we also calculate the number of the core spins as a function of time during the unzipping process [see Fig.~\ref{FIG14}(b)]. The number of the core spins is proportional to the total area of the spins pointing along the $-z$-direction, where the skyrmionium and/or skyrmions could be complete or broken (see Fig.~\ref{FIG12}). It can be seen that both the numbers of domain wall spins and core spins decrease with time. As pointed above, the exchange and PMA energies are proportional to the total length of the domain walls in the device. Indeed, we find the profile of the domain wall spins matches well with the profiles of the exchange and PMA energies with respect to time [compare Fig.~\ref{FIG13}(e) and Fig.~\ref{FIG14}(a)]. Also, we find the horizontally reversed profile of core spins agrees with the profile of the out-of-plane magnetization component with respect to time [compare Fig.~\ref{FIG13}(c) and Fig.~\ref{FIG14}(b)]. The reason is that the out-of-plane magnetization component is proportional to the area of the domains which are pointing along the $+z$-direction, and is inversely proportional to the area of the domains which are pointing along the $-z$-direction. It is worth mentioning that the number of core spins of the skyrmionium is basically two times than that of the two skyrmions [see Fig.~\ref{FIG14}(b)]. It indicates that the radius of the skyrmion is much smaller than that of the skyrmionium in the given device.

In Fig.~\ref{FIG15}(a), we show the out-of-plan magnetization component as a function of the number of core spins. We also show the exchange energy and DMI energy as functions of the number of domain wall spins in Fig.~\ref{FIG15}(b) and Fig.~\ref{FIG15}(c), respectively. Obviously, it can be seen that the out-of-plane magnetization component is inversely proportional to the number of core spins, that is, the area of the domains pointing along the $-z$-direction. On the other hand, it shows that the exchange energy and the DMI energy are proportional and antiproportional to the number of domain wall spins, that is the the total domain wall length in the device. Since the exchange interaction and the DMI prefer uniform and twisted spin structures, respectively, it is reasonable that larger domain wall spins, which are twisted spin structures, result in higher exchange energy and lower DMI energy.

\section{Conclusions}
\label{se:Conclusions}

In conclusion, we have investigated the generation of a skyrmionium in the nanodisk by applying the spin-polarized current with the CPP geometry. We have shown that the skyrmionium in the nanodisk can be controlled and manipulated by applying the perpendicular magnetic field, which could result in the degeneration of the skyrmionium with the topological transition of $Q=0\rightarrow Q=\pm 1$. We have also demonstrated that it is possible to drive the skyrmionium into motion in the nanotrack by applying the spin-polarized current, either with the CPP geometry or CIP geometry. It is found that the driving current with the CPP geometry can result in a better mobility of the skyrmionium. The steady velocity difference between the skyrmionium and skyrmion is proportional to the driving current density when they are driven by the spin-polarized current with the CPP geometry in the nanotrack. While the skyrmionium and skyrmion have no steady velocity difference when they are driven by the spin-polarized current with the CIP geometry in the nanotrack. Indeed, the distortion of the skyrmionium is found when the skyrmionium travels in the nanotrack, which is caused by the strong intrinsic SkHE. Importantly, we find that the distortion of the skyrmionium driven by the spin-polarized current with the CPP geometry is much more significant than that driven by the spin-polarized current with the CIP geometry for the same current density. Nevertheless, intriguingly there are parameter regions where the skyrmionium travels straightforward without any deformation or obvious deformation. We have also demonstrated that the bilayer-skyrmionium will not be distorted in the high-speed motion. In addition, we have shown that it is possible to transform a skyrmionium into a pair of skyrmions with the topological transition of $Q=0\rightarrow Q=+2$. Our results of the generation of a skyrmionium, its field-driven manipulation, as well as its current-driven transportation will be useful for designing future skyrmion-skyrmionium hybrid applications.

\begin{acknowledgments}
X.Z. was supported by JSPS RONPAKU (Dissertation Ph.D.) Program.
Y.Z. was supported by Shenzhen Fundamental Research Fund under Grant No. JCYJ20160331164412545.
W.S.Z. acknowledges the support by the projects from the Chinese Postdoctoral Science Foundation (No. 2015M570024), National Natural Science Foundation of China (Projects No. 61501013, No. 61471015 and No. 61571023), Beijing Municipal Commission of Science and Technology (Grant No. D15110300320000), and the International Collaboration Project (No. 2015DFE12880) from the Ministry of Science and Technology of China.
M.E. acknowledges the support by the Grants-in-Aid for Scientific Research from the JSPS KAKENHI (No. JP25400317 and No. JP15H05854).
\end{acknowledgments}



\end{document}